\documentclass[preprint,amsmath,amssymb,aps,prd,nofootinbib]{revtex4}
\usepackage{epsfig,graphicx,color,appendix}
\begin{document}
\begin{flushright}
CAU-THEP-19-03
\end{flushright}
\def\CP{{\it CP}~}
\def\cp{{\it CP}}
\title{\mbox{}\\[10pt]
Inflation and Leptogenesis \\in a $U(1)$-enhanced supersymmetric model}

\author{Y. H. Ahn}
\affiliation{Department of Physics, Chung-Ang University, Seoul 06974, Korea.}
\email{axionahn@naver.com}




\begin{abstract}
\noindent Motivated by the flavored Peccei-Quinn symmetry for unifying flavor physics and string theory, we investigate a supersymmetric extension of standard model (SM) for an explanation of inflation and leptogenesis by introducing $U(1)$ symmetries such that the $U(1)$-$[gravity]^2$ anomaly-free condition together with the SM flavor structure demands additional sterile neutrinos as well as no axionic domain-wall problem. Such additional neutrinos may play a crucial role as a bridge between leptogenesis and new neutrino oscillations along with high energy cosmic events. 
In a realistic moduli stabilization, we show that the moduli backreaction effect on the inflationary potential leads to the energy scale of inflation with the inflaton mass in a way that the power spectrum of the curvature perturbation and the scalar spectral index are to be well fitted with the latest Planck observation. We suggest that a new leptogenesis scenario could naturally be implemented via Affleck-Dine mechanism. So we show that the resultant baryon asymmetry, constrained by the sum of active neutrino masses and new high energy neutrino oscillations, crucially depends on the reheating temperature $T_{\rm reh}$. And we show that the model has a preference on $T_{\rm reh}\sim10^3$ TeV, which is compatible with the required $T_{\rm reh}$ to explain the baryon asymmetry of the Universe.
\end{abstract}

\maketitle %
\section{Introduction}
The standard model (SM) of particle physics has been successful in describing properties of known matter and forces to a great precision until now, but we are far from satisfied since it suffers from some problems or theoretical arguments that have not been solved yet, which follows: inclusion of gravity in gauge theory, instability of the Higgs potential, cosmological puzzles of matter-antimatter asymmetry, dark matter, dark energy, and inflation, and flavor puzzle associated with the SM fermion mass hierarchies, their mixing patterns with the CP violating phases, and the strong CP problem. The SM therefore cannot be the final answer. It is widely believed that the SM should be extended to a more fundamental underlying theory. If nature is stringy, string theory should give insight into all such fundamental problems or theoretical arguments\,\footnote{In Ref.\,\cite{Ahn:2016typ} a concrete model is designed to bridge between string theory as a fundamental theory and low energy flavor physics.}. As indicated in Refs.\,\cite{Ahn:2016typ,Ahn:2014gva}\,\footnote{Ref.\,\cite{Ahn:2014gva} introduces a superpotential for unifying flavor and strong CP problems, the so-called flavored PQ symmetry model in a way that no axionic domain wall problem.}, such several fundamental challenges strongly hint that a supersymmetric hybrid inflation framework with new gauge symmetries as well as higher dimensional operators responsible for the SM flavor puzzles may be a promising way to proceed.

Since astrophysical and cosmological observations have increasingly placed tight constraints on parameters for axion, neutrino, and inflation including the amount of reheating, it is in time for a new scenario on axion and neutrino to mount such interesting challenges, see also Ref.\,\cite{Ahn:2014gva, Ahn:2016hbn}.
In a theoretical point of view axion physics including neutrino physics requires new gauge interactions and a set of new fields that are SM singlets. Thus in extensions of the SM, sterile neutrinos and axions could naturally be introduced, {\it e.g.}, in view of $U(1)$ symmetry.
As a new paradigm to explain the aforementioned fundamental challenges, in this paper we investigate a minimal and economic supersymmetric extension of SM for an explanation of inflation and leptogenesis, which can be realized within the framework\,\footnote{Here the flavored Peccei-Quinn (PQ) symmetry $U(1)_X$ embedded in the non-Abelian $A_4$ finite group\,\cite{Ma:2001dn} could economically explain the mass hierarchies of quarks and leptons including their peculiar mixing patterns as well as provide a neat solution to the strong CP problem and its resulting axion\,\cite{Ahn:2016hbn}.} of $G\equiv SM\times U(1)_X\times A_4$. All renormalizable and nonrenormalizable operators allowed by such gauge symmetries, non-Abelian discrete symmetry, and $R$-parity exist in the superpotential as in Ref.\,\cite{Ahn:2016hbn}. Since non-perturbative quantum gravitational effects spoil the axion solution to the strong CP problem\,\cite{Kamionkowski:1992mf, Dvali:2005an}, in order to eliminate such breaking effects of the axionic shift symmetry by gravity the author in Ref.\,\cite{Ahn:2016hbn} has imposed an $U(1)_X\times[gravity]^2$ anomaly cancellation condition\,\cite{Ahn:2016hbn} in a way that no axionic domain-wall problem occurs, thereby additional sterile neutrinos are introduced. Such sterile neutrinos are light or heavy and do not participate in the weak interaction. Moreover, the latest results\,\cite{Planck2014} from Planck and Baryon Acoustic Oscillations (BAO) show that the contribution of light sterile neutrinos to $N^{\rm eff}_{\nu}$ at the Big-Bang Nucleosynthesis (BBN)\,\cite{BBN} era is negligible; such neutrinos may play a crucial role as a bridge between leptogenesis and new neutrino oscillations along with high energy cosmic events.

In this paper, in order to provide an explanation for inflation we present a realistic moduli stabilization, which is essential for the flavored PQ axions to be realized at low energy scale\,\cite{Ahn:2016hbn}.
Such moduli stabilization has moduli backreaction effects on the inflationary potential, 
which could provide a lucid explanation for the cosmological inflation at high energy scale.
 Thus such moduli stabilization with the moduli backreaction effects on the inflationary potential leads to the energy scale of inflation with the inflaton mass, $m_{\Psi_0}=\sqrt{3}\,H_I$, in a way that the power spectrum of the curvature perturbation and the scalar spectral index are to be well fitted with the latest Planck observation\,\cite{Ade:2015xua}.
And we suggest, interestingly enough, a new leptogenesis scenario which could naturally be implemented through Affleck-Dine (AD) mechanism for baryogenesis\,\cite{Affleck:1984fy} and its subsequent leptonic version so-called AD leptogenesis\,\cite{Murayama:1993em}.
 Interestingly enough, the pseudo-Dirac mass splittings, suggested from the new neutrino oscillations along with high energy cosmic events\,\cite{Ahn:2016hbn}, strongly indicate the existence of lepton-number violation which is a crucial ingredient of the present leptogenesis scenario.
 So the resultant baryon asymmetry is constrained by the cosmological observable ({\it i.e.} the sum of active neutrino masses) with the new high energy neutrino oscillations, and crucially depends on the reheating temperature which depends on gravitational and non-gravitational decays of the inflaton and waterfall field. Since all the particles including photons and baryons in the present universe are ultimately originated from the inflaton and waterfall field decays, it is crucial to reveal how the reheating proceeds. We show that the reheating temperature is mainly determined by the non-gravitational decay of the waterfall field, leading to a relatively low reheating temperature 
 which is consistent with that for explaining the right value of  the baryon asymmetry of the universe (BAU), $Y_{\Delta B}\simeq8\times10^{-11}$\,\cite{Ade:2015xua}, together with the pseudo-Dirac mass splittings responsible for new oscillations $\Delta m^2_i\simeq{\cal O}(10^{-12})$ eV$^2$.
In addition, since gravitinos are present in the supersymmetric model we are going to address gravitino overabundance problem.
We consider direct decays of the inflaton to gravitinos competing with the thermal production in the thermal plasma formed after reheating when setting limits on the couplings governing inflaton decay, see Eq.\,(\ref{toyi1}).

The rest of this paper is organized as follows. In Sec.\,II we setup and review the model based on $A_4\times U(1)_X$ symmetry in order to investigate an economic SUSY inflationary scenario and a new leptogenesis via AD mechanism.
In Sec.\,III, first we study a realistic moduli stabilization in type IIB string theory with positive vacuum energy, which is essential for the flavored PQ axions at low energy as well as a lucid explanation for cosmological inflation at high energy
scale.
And we investigate how the size moduli stabilized at a scale close to $\Lambda_{\rm GUT}$ significantly affect the dynamics of the inflation, as well as how the $X$-symmetry breaking scale during inflation is induced and its scale is fixed at $\sim0.3\times10^{16}$ GeV by the amplitude of the primordial curvature perturbation and the spectral index. 
The main focus on Sec.\,IV is to show that a successful leptogenesis scenario could be naturally implemented through AD mechanism, and subsequently estimate the reheating temperature that is required to generate sufficient lepton number asymmetry following the hybrid $F$-term inflation. In turn, we show that the successful leptogenesis is closely correlated with the neutrino oscillations available on high- and low-energy neutrinos, and how the amount of reheating could be strongly correlated with the successful leptogenesis.
Moreover, we discuss that it is reasonable for the reheating temperature $T_{\rm reh}\sim10^3$ TeV derived from the gravitational decays of the inflaton and waterfall field to be compatible with the required reheating temperature for the successful leptogenesis.
What we have done is summarized in Sec.\,V.

\section{flavor $A_{4}\times U(1)_{X}$ symmetry and setup}
\label{A4U1}
Unless flavor symmetries are assumed, particle masses and mixings are generally undetermined in the SM gauge theory. In order to provide an elegant solution to the strong CP problem and describe the present SM flavor puzzles associated with the fermion mass hierarchies including their mixing patterns, the author in Ref.\,\cite{Ahn:2014gva, Ahn:2016hbn} has introduced the non-Abelian discrete $A_{4}$ flavor symmetry\,\cite{Altarelli:2005yp, nonAbelian} which is mainly responsible for the peculiar mixing patterns, as well as an additional continuous symmetry $U(1)_{X}$ which is mainly for vacuum configuration as well as for  describing mass hierarchies of leptons and quarks.
In Ref.\,\cite{Ahn:2016hbn} the symmetry group for matter fields (leptons and quarks), flavon fields and driving fields\,\footnote{The flavon fields are responsible for the spontaneous breaking of the flavor symmetry, while the driving fields are introduced to break the flavor group along required vacuum expectation value (VEV) directions and to allow the flavons to get VEVs, which couple only to the flavons, see Appendix\,\ref{dri}.} is $A_{4}\times U(1)_{X}$ where $U(1)_X\equiv U(1)_{X_1}\times U(1)_{X_2}$. 
We take the $U(1)_{X_1}$ breaking scale corresponding to the $A_{4}$ symmetry breaking scale and the $U(1)_{X_2}$ breaking scale to be separated by Gibbons-Hawking temperature, $T_{\rm GH}=H_I/2\pi$, and both of which are to be much above the electroweak scale in our scenario\,\footnote{See the symmetry breaking scales from the astrophysical constraints\,\cite{Ahn:2016hbn}, and in more detail  Sec.\,\ref{inf_PQ1} on the PQ symmetry breaking scale during inflation.}, that is, 
\begin{eqnarray}
 \langle H_{u,d}\rangle\ll\langle\Phi_{T}\rangle,\langle\Phi_{1}\rangle<\frac{H_I}{2\pi}<\langle\Phi_2\rangle
 \label{hierarchy_vev}
\end{eqnarray}
where $H_{I}$ is the inflationary Hubble constant, and the fields $\Phi_1=\{\Phi_S, \Theta\}$ and $\Phi_2=\{\Psi, \tilde{\Psi}\}$ are charged under the $U(1)_{X_1}$ and $U(1)_{X_2}$ symmetries, respectively. 
So we can picture two secluded SUSY breaking sectors by the inflationary sector and by the visible sector in the present Universe, {\it i.e.}, SUSY$=$SUSY$_{\rm inf}\times$SUSY$_{\rm vis}$, respectively. Both sectors interact non-gravitationally via inflaton field as well as gravitationally. Since the Kahler moduli superfields putting the GS mechanism into practice are not separated from the SUSY$_{\rm inf}$ during inflation, the $U(1)_{X_2}$-charged matter fields develop a large VEV during inflation by taking tachyonic SUSY breaking scalar masses $m^2_{\Phi_2}\sim-H^2_I$ induced `dominantly' by the $U(1)_{X_2}$ $D$-term, compared to the Hubble induced soft masses generated by the $F$-term SUSY breaking. On the other hand, in the present Universe both the $U(1)_{X_i}$-charged matter fields $\Phi_{1}$ and $\Phi_{2}$ develop large VEVs by the soft-SUSY breaking mass. So, in the absence of direct interactions, gravitational or otherwise, the $U(1)_{X_2}$-charged chiral superfields $\Phi_2$ have a two-fold enhanced SUSY$_{\rm inf}\times$SUSY$_{\rm vis}$ Poincare symmetry. However, gravitational interactions explicitly break the SUSY down to {\it true} SUSY$_{\rm inf}\times$SUSY$_{\rm vis}$, where SUSY$_{\rm inf}$ corresponds to the genuine SUGRA symmetry, while the orthogonal SUSY$_{\rm vis}$ is only approximate global symmetry. In each sector, spontaneous breakdown of $F$-term occurs at a scale $F_i$ ($i=$ inf, vis) independently, producing a corresponding goldstino. In the presence of SUGRA, SUSY$_{\rm inf}$ is gauged and thus its corresponding goldstino is eaten by the gravitino via super-Higgs mechanism, leaving behind the approximate global symmetry SUSY$_{\rm vis}$ which is explicitly broken by SUGRA and thus its corresponding the uneaten goldstino as a physical degree of freedom.
During inflation and the beginning of reheating (preheating) the SUSY$_{\rm inf}$ is mainly broken by the inflaton implying the goldstino produced is mainly inflatino; the gravitino produced non-thermally is effectively massless as long as $H>m_{3/2}$. However, this correspondence does not necessarily hold at late times, since the SUSY$_{\rm vis}$ is broken by other field in the true  vacuum implying that the corresponding uneaten goldstino gives masses mainly to all the supersymmetric SM superpartners in the visible sector.

\section{Inflation}
 \label{inflat}
The inflation that inflated the observable universe beyond the Hubble radius, and could have produced the seed inhomogeneities needed for galaxy formation and the anisotropies observed by COBE\,\cite{cobe}, must occur at an energy scale $V^{1/4}\leq4\times10^{16}$ GeV\,\cite{Liddle:1993ch}, well below the Planck scale. At this relatively low energies, superstrings are described by an effective ${\cal N} = 1$ supergravity theory\,\cite{Nilles:1983ge}. 
We work in the context of supersymmetric moduli stabilization, in the sense that all moduli masses are independent of the gravitino mass and large compared to the scale of any other dynamics in the effective theory, e.g., the scale of inflation, $m_{T_{i}}>H_I$ where $H_I=\sqrt{V/3M^2_P}$ is the Hubble scale during inflation. As in Ref\,\cite{Ahn:2016typ, Ahn:2016hbn}, the size moduli with positive masses have been stabilized, while leaving two axions massless and one axion massive, {\it i.e.} $m_T\sim m_{\theta^{\rm st}}\gg m_{3/2}$. So we will discuss that such moduli stabilization has moduli backreaction effects on the inflationary potential, in particular, the spectral index of inflaton fluctuations, which provides a lucid explanation for the cosmological inflation at high energy scale. We are going to see how the size moduli stabilized at a scale close to $\Lambda_{\rm GUT}$ significantly affect the dynamics of the inflation, as well as how the $X$-symmetry breaking scale during inflation is induced and its scale is fixed at $\sim0.7\times10^{16}$ GeV, close to $\Lambda_{\rm GUT}$, by the amplitude of the primordial curvature perturbation.

The model addressed in Refs.\,\cite{Ahn:2014gva, Ahn:2016typ} naturally causes a hybrid inflation\,\footnote{Supersymmetirc realizations of $F$-term hybrid inflation were first studied in Ref.\,\cite{Copeland:1994vg}. And the hybrid inflation model in supergravity\,\cite{elec_vev, Lazarides:1996dv, Senoguz:2004vu} and the $F$-term hybrid inflation in supersymmetric moduli stabilization~\cite{Kallosh:2010xz} were studied in detail. See also Refs.\,\cite{D_term, Garbrecht:2006az}}, in which the QCD axion and the lightest neutralino charged under a stabilizing symmetry could become components of dark mater.
We work in a SUGRA framework based on type IIB string theory, and assume that the dilaton and complex structure moduli are fixed at semi-classical level by turning on background fluxes\,\cite{Gukov:1999ya}. 
Below the scale where the complex structure and the axio-dilaton moduli are stabilized through fluxes as in Refs.\,\cite{Giddings:2001yu, Dasgupta:1999ss}, 
in Einstein frame\,\footnote{In Jordan frame since the sign of the kinetic term for the scalar field is not positive definite one could not have a stable ground state. Hence the correct procedure is to transform the potentials to the Einstein frame, and then the system in Einstein frame could not decay lower energy states\,\cite{Magnano:1993bd}. And with this correct procedure we found a blot on this work that
the procedure break down for large couplings of the inflaton to the Ricci scalar.} the SUGRA scalar potential is
\begin{eqnarray}
 V= e^{G}M^4_P\Big(\sum_\alpha G^\alpha G_\alpha-3\Big)+\frac{1}{2}f^{-1}_{ij}D^iD^j\,,
 \label{scapot}
\end{eqnarray}
where $G^\alpha=G^{\alpha\bar{\beta}}G_{\bar{\beta}}$ with $G^{\alpha\bar{\beta}}=M^2_PK^{\alpha\bar{\beta}}$, $M_{P}=(8\pi G_N)^{-1/2}=2.436\times10^{18}$ GeV is the reduced Planck mass with the Newton's gravitational constant $G_N$, and $f_{ij}$ is the gauge kinetic function. And the $F$-term potential is given by the first term in the right hand side of Eq.\,(\ref{scapot}); the $D$-term, the second term in the right hand side of Eq.\,(\ref{scapot}), is quartic in the charged fields under the gauge group, and in the model it is flat along the inflationary trajectory so that it can be ignored during inflation\,\footnote{Assuming the FI $D$-terms do not appear during inflation, $\xi^{\rm FI}_i=0$, it is likely that $D$ terms in the inflaton sector do not give a significant contribution to the inflaton potential. See Sec.\,\ref{inf_PQ1}.}. The generalized Kahler potential, $G$, is given by
\begin{eqnarray}
 G=\frac{K}{M^2_P}+\ln\frac{|W|^2}{M^6_P}\,.
\end{eqnarray}
Here the low-energy Kahler potential $K$ and superpotential $W$ for moduli and matter superfields, invariant under $U(1)_X$ gauged symmetry, are given in type IIB string theory by\,\cite{Ahn:2016typ}
 \begin{eqnarray}
  K&=&-M^2_P\ln\Big\{(T+\bar{T})\prod^2_{i=1}\big(T_i+\bar{T}_i-\frac{\delta^{\rm GS}_i}{16\pi^2}V_{X_i}\big)\Big\}+\tilde{K}+...\label{Kahler0}\\
  &&\qquad\text{with}\quad\tilde{K}=\sum^2_{i=1}Z_i\Phi^\dag_i e^{-X_iV_{X_i}}\Phi_i+\sum_k Z_k|\varphi_k|^2\,,\nonumber\\
  W&=& W_{Y}+W_v+W_0+W(T)\label{Kahler}\,,
 \label{KaMo}
 \end{eqnarray}
in which $\Phi_1=\{\Phi_S, \Theta,\tilde{\Theta}\}$, $\Phi_2=\{\Psi,\tilde{\Psi}\}$,  $\varphi_{i}=\{\Psi_0, \Phi^{T}_{0},\Phi_{T}\}$,  dots represent higher-order terms. $W_0$ stands for the constant value of the flux superpotential at its minimum. Since the Kahler moduli do not appear in the superpotential $W$ at leading order, they are not fixed by the fluxes. So a non-perturbative superpotential $W(T)$ is introduced to stabilize the Kahler moduli\,\cite{Ahn:2016typ}, although $W(T)$ in Eq.\,(\ref{Kahler}) is absent at tree level. 
The Kahler moduli in $K$ of Eq.~(\ref{Kahler0}) control the overall size of the compact space, 
 \begin{eqnarray}
  T=\rho+i\theta,\qquad T_i=\rho_i+i\theta_i \quad\text{with}~i=1,2\,,
 \end{eqnarray}
where $\rho(\rho_i)$ are the size moduli of the internal manifold and $\theta(\theta_i)$ are the axionic parts.
As can be seen from the Kahler potential above, the relevant fields participating in the four-dimensional Green-Schwarz (GS) mechanism\,\cite{GS} are the $U(1)_{X_i}$ charged chiral matter superfields $\Phi_i$, the vector superfields $V_{X_i}$ of the gauged $U(1)_{X_i}$ which is anomalous, and the Kahler moduli $T_i$. The matter superfields in $K$ consist of all the scalar fields $\Phi_i$ that are not moduli and do not have Planck sized VEVs, and the chiral matter fields $\varphi_k$ are neutral under the $U(1)_{X_i}$ symmetry. We take, for simplicity, the normalization factors $Z_i=Z_k=1$, and the holomorphic gauge kinetic function $f_{ij}=\delta_{ij} (1/g^2_j+ia_{T_j}/8\pi^2)$, {\it i.e.}, $T_i=1/g^2_{X_i}+ia_{T_i}/8\pi^2$ on the Kahler moduli in the 4-dimensional effective SUGRA where $g_{X_i}$ are the four-dimensional gauge couplings of $U(1)_{X_i}$. Actually, gaugino masses require a nontrivial dependence of the holomorphic gauge kinetic function on the Kahler moduli. This dependence is generic in most of the models of ${\cal N}=1$ SUGRA derived from extended supergravity and string theory\,\cite{Ferrara:2011dz}. And vector multiplets $V_{X_i}$ in Eq.\,(\ref{Kahler0}) are the $U(1)_{X_i}$ gauge superfields including gauge bosons $A^{\mu}_i$. The GS parameter $\delta^{\rm GS}_i$ characterizes the coupling of the anomalous gauge boson to the axion.

Non-minimal SUSY hybrid inflation can be defined by the superpotential $W_{\rm inf}$ which is an analytic function, together with  a Kahler potential $K_{\rm inf}$ which is a real function
  \begin{eqnarray}
  &W\supset W_{\rm inf}=g_7\,\Psi_0\left(\Psi\tilde{\Psi}-\mu^2_\Psi\right)\,,\label{NK0}\\
  &\tilde{K}\supset K_{\rm inf}=|\Psi_0|^2+|\Psi|^2+|\tilde{\Psi}|^2+k_{s}\frac{|\Psi_0|^4}{4M^2_{P}}+k_{1}\frac{|\Psi_0|^2|\Psi|^2}{M^2_{P}}+k_{2}\frac{|\Psi_0|^2|\tilde{\Psi}|^2}{M^2_{P}}+k_{3}\frac{|\Psi_0|^6}{6M^4_{P}}+...
  \label{NK}
 \end{eqnarray}
where $\Psi_0$ and $\Psi(\tilde{\Psi})$ denote the inflaton and PQ fields, respectively. Here the dimensionless couplings $g_7, k_s, k_{1,2,...}$ are of order unity. The PQ scalar fields play a role of the waterfall fields, that is, the PQ phase transition takes place during inflation such that the PQ scale $\mu_\Psi=\mu_\Psi(t_I)$ sets the energy scale during inflation.

The kinetic terms of the Kahler moduli and scalar sectors in the flat space limit of the 4 dimensional ${\cal N} = 1$ supergravity are expressed as
  \begin{eqnarray}
   {\cal L}_{\rm kinetic}=K_{T\bar{T}}\,\partial_\mu{T}\partial^{\mu}\bar{T}+K_{T_i\bar{T}_i}\,\partial_\mu{T}_i\partial^{\mu}\bar{T}_i+K_{\Phi_i\bar{\Phi}_i}\,\partial_\mu{\Phi}_i\partial^{\mu}{\Phi}^\dag_i\,.
  \label{kinetic}
 \end{eqnarray}
Here we set $K_{\Phi_i\bar{\Phi}_i}=1$ for canonically normalized scalar fields.
In addition to the superpotential in Eq.\,(\ref{Kahler}) the Kahler potential in Eq.\,(\ref{Kahler0}) deviates from the canonical form due to the contributions of non-renormalizable terms scaled by an UV cutoff $M_P$, invariant under the both gauge and the flavor symmetries.
\subsection{Supersymmetric Moduli Stabilization}
In string theory, one must consider stabilization of the volume moduli to explain why our universe is 4-dimensional rather than 10-dimensional. 
Since the three moduli all appear in the Kahler potential Eq.\,(\ref{Kahler0}), by solving the $F$-term equations the three size moduli and one axionic partner with positive masses are stabilized while leaving two axions massless through an effective superpotential $W(T)$\,\cite{Ahn:2016typ}. As will be seen later, the two massless axion directions will be gauged by the $U(1)$ gauge interactions associated with $D$-branes, and the gauged flat directions of the $F$-term potential will be removed through the Stuckelberg mechanism. The $F$-term scalar potential has the form 
 \begin{eqnarray}
   V_F=\frac{e^{\tilde{K}/M^2_P}}{(T+\bar{T})(T_1+\bar{T}_1)(T_2+\bar{T}_2)}\Big\{\sum_{I=T,T_1,T_2}K^{I\bar{I}}|D_I W|^2-\frac{3}{M^2_P}|W|^2+K^{i\bar{i}}|D_{i} W|^2\Big\}
 \label{}
 \end{eqnarray}
for $V_{X_i}=0$, where $K^{I\bar{J}}=0$ for $I\neq J$, and $I,J$ stand for $T,T_i$ and $i,j$ for the bosonic components of the superfields $\Phi_i,\varphi_i$. 
Here the Kahler covariant derivative and Kahler metric are defined as $D_IW\equiv\partial_I W+W\partial_I K/M^2_P$ and  $K_{I\bar{J}}\equiv\partial_I\partial_{\bar{J}}K$,
where $D_{\bar{I}}\overline{W}=(\overline{D_I} \overline{W})$, and $K^{I\bar{J}}$ is the inverse Kahler metric $(K)^{-1}_{I\bar{J}}$.
In order for the Kahler moduli $T$ and $T_i$ to be stabilized certain non-perturbative terms are introduced as an effective superpotential\,\cite{Ahn:2016typ}
\begin{eqnarray}
 W(T)&=&A(\Phi_i)e^{-a(T+T_1+T_2)}+B(\Phi_i)e^{-b(T+T_1+T_2)}\,,
 \label{stabilization}
\end{eqnarray}
where the coefficients $a=2\pi$ or $2\pi/N$ and $b=2\pi$ or $2\pi/M$ are the corrections arising from $D3$ instantons or gaugino condensation in a theory with a product of non-Abelian gauge groups $SU(N)\times SU(M)$. 
Here $A(\Phi_i)$ and $B(\Phi_i)$ are analytic functions of $\Phi_i$ transforming under $U(1)_{X_i}$ as 
 \begin{eqnarray}
  A(\Phi_i)\rightarrow A(\Phi_i)\,e^{i\frac{a}{16\pi^2}(\delta^{\rm GS}_1\Lambda_1+\delta^{\rm GS}_2\Lambda_2)}\,,\qquad
  B(\Phi_i)\rightarrow B(\Phi_i)\,e^{i\frac{b}{16\pi^2}(\delta^{\rm GS}_1\Lambda_1+\delta^{\rm GS}_2\Lambda_2)}\,,
 \label{WT_trns}
 \end{eqnarray}
 and invariant under the other gauge group. Since there are two non-perturbative superpotentials of the form $W_{\rm np}=Ae^{-aT}$,  the structure of the effective scalar potential has two non-trivial minima at different values of finite $T_{(i)}$. One corresponds to a supersymmetric Minkowski vacuum which could be done through the background fluxes $W_0$, while the other corresponds to a negative cosmological constant which gives rise to a supersymmetric Anti de Sitter (AdS) vacuum.
So the height of the barrier separates the local Minkowski minimum from the global AdS minimum, and the gravitino mass vanishes at the supersymmetric Minkowski minimum. As will be seen in Eq.\,(\ref{inflaton_mass}), inflaton mass ($m_{\Psi_0}\sim H_I$) is much smaller than the size moduli masses, and consequently the size moduli will be frozen quickly during inflation without perturbing the inflation dynamics. And it is expected that $H_I\ll\Lambda_{\rm GUT}$ as a consequence of the enormous flatness of inflaton potential, where $\Lambda_{\rm GUT}\simeq2\times10^{16}$ GeV is the scale gauge coupling unification in the supersymmetric SM.
 The scalar potential of the fields $\rho$ and $\rho_{i}$ has local minimum at $\sigma_0,\sigma_i$ which is supersymmetric, i.e., 
 \begin{eqnarray}
  W(\sigma_0,\sigma_i)=0\,,\qquad D_TW(\sigma_0,\sigma_i)=D_{T_i}W(\sigma_0,\sigma_i)=0\,,
 \label{}
 \end{eqnarray}
 and Minkowski, i.e.,
  \begin{eqnarray}
  V_F(\sigma_0,\sigma_i)=0\,,
 \label{vf0}
 \end{eqnarray}
 where  $\sigma_0=\sigma_i=\frac{1}{a-b}\ln\left(\frac{a\,A_0}{b\,B_0}\right)$. And $W_0$ is fine-tuned as 
 \begin{eqnarray}
  W_0=-A_0\left(\frac{a\,A_0}{b\,B_0}\right)^{-3\frac{a}{a-b}}-B_0\left(\frac{a\,A_0}{b\,B_0}\right)^{-3\frac{b}{a-b}}\,,
 \label{W0}
 \end{eqnarray}
 where $A_0$ and $B_0$ are constant values of order ${\cal O}(1)$ of $A(\Phi_i)$ and $B(\Phi_i)$, respectively, at a set of VEVs $\langle\Phi_i\rangle$ that cancel all the $D$-terms, including the anomalous $U(1)_{X_i}$, see Ref.\,\cite{Ahn:2016hbn}. Here the constant $W_0$ is not analytic at the VEVs $\langle\Phi_i\rangle$, where the moduli are stabilized at the local supersymmetric Minkowski minumum. Moreover, since $W(T)$ is an effective superpotential its analyticity does not need to be guaranteed in the whole range of the $\Phi_i$ fields, and so, as will be shown later, the anomalous FI terms at the global supersymmetric AdS minimum can not be cancelled and act as uplifting potentials. Restoration of supersymmetry in the supersymmetric local Minkowski minimum implies that all particles whose mass is protected by supersymmetry are expected to light in the vicinity of the minimum. However, supersymmetry breaks down and all of these particles become heavy once one moves away from the minimum of the effective potential. This is exactly the situation required for the moduli trapping near the enhanced symmetry points\,\cite{Kofman:2004yc}.

The $F$-term equations $D_TW=D_{T_i}W=0$, where we set the matter fields to zero, provide $\rho=\rho_i$, and lead to
 \begin{eqnarray}
  aA\,e^{-3a\rho}\,e^{-ia\,\theta^{\rm st}}+bB\,e^{-3b\rho}\,e^{-ib\,\theta^{\rm st}}
  +\frac{W_0+A\,e^{-3a\rho}\,e^{-ia\,\theta^{\rm st}}+B\,e^{-3b\rho}\,e^{-ib\,\theta^{\rm st}}}{2\rho}=0
 \label{stabil}
 \end{eqnarray}
for $V_{X_i}=0$, where $\theta^{\rm st}\equiv\theta+\theta_1+\theta_2$. This shows that the three size moduli $(\rho,\rho_i)$ and one axionic direction $\theta^{\rm st}$ are fixed, while the other two axionic directions ($\theta^{\rm st}_1\equiv\theta-\theta_1$ and $\theta^{\rm st}_2\equiv\theta-\theta_2$) are independent of the above equation. So, without loss of generality, we rebase the superfields $T$ with $\theta^{\rm st}={\rm Im}[T]$ and $T_i$ with $\theta^{\rm st}_i={\rm Im}[T_i]$ as
 \begin{eqnarray}
   T=\rho+i\theta&\rightarrow& T=\rho+i\theta^{\rm st}\,,\nonumber\\
   T_{i}=\rho_i+i\theta_{i}&\rightarrow& T_{i}=\rho_i+i\theta^{\rm st}_{i}\,.
 \label{redf_K}
 \end{eqnarray}
Then  from the $F$-term scalar potential, while the gravitino mass in the supersymmetric local Minkowski minimum vanishes, the masses of the fields $\rho$, $\rho_{1}$, $\rho_{2}$, and $\theta^{\rm st}$, respectively, are obtained as
 \begin{eqnarray}
   m^{2}_{T}&=&\frac{1}{2}K^{T\bar{T}}\partial_T\partial_{\bar{T}}V_F\Big|_{T=\bar{T}=\sigma_0}\nonumber\\
   &=&\frac{3\ln\Big(\frac{a\,A_0}{b\,B_0}\Big)}{M^4_P(a-b)}\Big\{A_0\,a^2\Big(\frac{a\,A_0}{b\,B_0}\Big)^{-3\frac{a}{a-b}}+B_0\,b^2\Big(\frac{a\,A_0}{b\,B_0}\Big)^{-3\frac{b}{a-b}}\Big\}^2\,,\nonumber
 \end{eqnarray}
 \begin{eqnarray}    
 m^{2}_{\theta^{\rm st}}&=&\frac{1}{2}K^{T\bar{T}}\partial_{\theta^{\rm st}}\partial_{\theta^{\rm st}}V_F\Big|_{T=\bar{T}=\sigma_0}\nonumber\\
 &=&\frac{3W_0}{M^4_P}\Big\{-A_0\,a^3\Big(\frac{a\,A_0}{b\,B_0}\Big)^{-3\frac{a}{a-b}}-B_0\,b^3\Big(\frac{a\,A_0}{b\,B_0}\Big)^{-3\frac{b}{a-b}}\Big\}\nonumber\\
 &+&\frac{6\ln\left(\frac{a\,A_0}{b\,B_0}\right)}{M^4_P(a-b)}\Big\{-A_0\,B_0(a-b)^2\Big(\frac{a\,A_0}{b\,B_0}\Big)^{-3\frac{a+b}{a-b}}\Big(\frac{a^2-b^2}{2\ln\big(\frac{a\,A_0}{b\,B_0}\big)}+a\,b\Big)\Big\}\,,
 \label{massT}
 \end{eqnarray}
where $a,b$ are positive constants, while $A_0,B_0$ are constants in $M^3_P$ units. Here the mass squared of the size moduli fields $\rho_{(i)}$ at the minimum is given by 
  $m^2_T\equiv m^{2}_{\rho}=m^{2}_{\rho_{i}}=3\,\sigma_0\,\left|W_{TT}(\sigma_0)\right|^2/M^4_P$
where $W_{TT}|_{\rm all\,matter\,fields=0}=a^2\,A\,e^{-a(T+T_1+T_2)}+b^2\,B\,e^{-b(T+T_1+T_2)}$ with $W_{TT}\equiv\partial^2W/(\partial T)^2$. 
 With the conditions $a>0$ and $b>0$ we obtain positive values of masses: as an example, for $A_0=-2.13$ and $B_0=-1.65$ with inputs $a=2\pi/100$ and $b=2\pi/60$, we obtain $\sigma_0\simeq6.17$, $W_0\simeq-0.90$ and\,\footnote{These values ensure $m_T\sim10^{16-17}$ GeV and $|\tilde{g}_7|={\cal O}(1)\times10^{-3}$ through $\tilde{g}^2_7=g^2_7/(2\sigma_0)^3$ in Eq.\,(\ref{moduli_V}), satisfying the two observables, {\it i.e.}, the scalar spectral index $n_s$ and the power spectrum of the curvature perturbations $\Delta^2_{\cal R}(k_0)$ in TABLE\,\ref{inf_result}.}
 \begin{eqnarray}
  m_T\simeq5.47\times10^{16}\,{\rm GeV}\,\qquad m_{\theta_{\rm st}}\simeq7.61\times10^{16}\,{\rm GeV}\,,
 \label{m_TT}
 \end{eqnarray}
 numerically.
 Note that due to the relation $(aA_0/bB_0)^{\frac{1}{a-b}}=e^{\sigma_0}$, see below Eq.\,(\ref{vf0}), as the masses $m_T$ and $m_{\theta^{\rm st}}$ increase the value of $\sigma_0$ decreases.
As will be seen in Sec.\,\ref{inflat} and in TABLE\,\ref{inf_result}, the moduli stabilized at a scale close to $\Lambda_{\rm GUT}$ will significantly affect the dynamics of the inflation and well fit the cosmological observables.
 
\subsection{Supersymmetry breaking and Cosmological constant}
\label{susy_break}
As discussed before, the supersymmetric local Minkowski vacuum at $\rho=\sigma_0$ and $\rho_i=\sigma_i$ is absolutely stable with respect to the tunneling to the vacuum with a negative cosmological constant because the Minkowski minimum is separated from a global AdS minimum by a high barrier. This vacuum state becomes metastable after uplifting of a AdS minimum to the dS minimum with $\Lambda_c\sim10^{-120}\,M^4_P$.
The other supersymmetric global AdS minimum is defined by 
 \begin{eqnarray}
  W(\sigma_{\tilde{0}},\sigma_{\tilde{i}})\neq0\,\qquad D_TW(\sigma_{\tilde{0}}, \sigma_{\tilde{i}})=D_{T_i}W(\sigma_{\tilde{0}},\sigma_{\tilde{i}})=0\,,
 \label{}
 \end{eqnarray}
corresponding to the minimum of the potential with $V_{\rm AdS}<0$. And at this AdS minimum one can set the value of the superpotential $\Delta W\equiv\langle W\rangle_{\rm AdS}$ by tuning $W_0$ at values of finite $\sigma_{\tilde{0}},\sigma_{\tilde{i}}$.
The existence of FI terms $\xi^{\rm FI}_i$ for the corresponding $U(1)_{X_i}$ implies the existence of uplifting potential which makes a nearly vanishing cosmological constant and induces SUSY breaking. A small perturbation $\Delta W$ to the superpotential\,\cite{Kallosh:2004yh, Kallosh:2003} is introduced in order to determine SUSY breaking scale. Then the minimum of the potential is shifted from zero to a slightly negative value at $\sigma_{\tilde{0}}=\sigma_0+\delta\rho$ and $\sigma_{\tilde{i}}=\sigma_i+\delta\rho_i$ by the small constant $\Delta W$. The resulting $F$-term potential has a supersymmetric AdS minimum and consequently the depth of this minimum is given by $V_{\rm AdS}=-3\,e^{\tilde{K}/M^2_P}\frac{|W|^2}{M^2_P}$; which can be approximated in terms of $W(\sigma_0+\delta\rho, \sigma_i+\delta\rho_i)\simeq\Delta W+{\cal O}(\Delta W)^2$ as
 \begin{eqnarray}
  V_{\rm AdS}(\Delta W)\simeq-\frac{3}{M^2_P}\frac{(\Delta W)^2}{8\sigma_0\sigma_1\sigma_2}=-\frac{3}{8M^2_P}\Big(\frac{a-b}{\ln\frac{aA_0}{bB_0}}\Big)^2(\Delta W)^2\,.
 \label{AdS}
 \end{eqnarray}
At the shifted minimum SUSY is preserved, {\it i.e.} $D_TW(\sigma_0+\delta\rho)=0$ and $D_{T_i}W(\sigma_i+\delta\rho_i)=0$, leading to $W_T(\sigma_0+\delta\rho)=W_{T_i}(\sigma_0+\delta\rho_i)\simeq3\Delta W/2\sigma_0$.
At this new minimum the displacements $\delta\rho=\delta\rho_{i}$ are obtained as
 \begin{eqnarray}
  \delta\rho_{(i)}\simeq\frac{3\Delta W}{2\sigma_0 W_{TT}(\sigma_0)}
  = \frac{3(a-b)\Delta W}{2\ln\left(\frac{aA_0}{bB_0}\right)\Big\{A_0a^2\Big(\frac{aA_0}{bB_0}\Big)^{\frac{-3a}{a-b}}+B_0b^2\Big(\frac{aA_0}{bB_0}\Big)^{\frac{-3b}{a-b}}\Big\}}\,.
 \label{}
 \end{eqnarray}
After adding the uplifting potentials SUSY is broken and then the gravitino in the uplifted minimum acquires a mass $ m^2_{3/2}=\langle e^{\tilde{K}/M^2_P}\rangle\,|W|^2/M^4_P$ :
 \begin{eqnarray}
  m_{3/2}=\sqrt{\frac{|V_{\rm AdS}|}{3M^2_P}}\simeq\frac{|\Delta W|}{M^2_P}\Big(\frac{a-b}{2\ln\frac{aA_0}{bB_0}}\Big)^{\frac{3}{2}}\,.
 \label{gravitino}
 \end{eqnarray}
The important point is that the masses $m_{T}$ and $m_{\theta^{\rm st}}$ in Eq.\,(\ref{massT}), as well as the height of barrier from the runaway direction,  do not have any relation to the gravitino mass, {\it i.e.}, $m_{T}\sim m_{\theta^{\rm st}}\gg m_{3/2}$. Thus we will consider the $F$-term hybrid inflation for $H_I\gg m_{3/2}$ in the Sec.~\ref{inflat}.

The uplifting of the AdS minimum to the dS minimum can be achieved by considering non-trivial fluxes for the gauge fields living on the D7 branes\,\cite{Burgess:2003ic} which can be identified as field-dependent FI $D$-terms in the ${\cal N}=1$, $4D$ effective action\,\cite{Brunner:1999jq}. As shown in Refs.\,\cite{Burgess:2003ic}, uplifting of the AdS minimum induces SUSY breaking and is achieved by adding to the potential two terms $\Delta V_i\approx|V_{\rm AdS}|\sigma^3_{\tilde{i}}/\rho^3$ if the uplifting term occurs due to a $D$-term.
Similarly, we can parameterize the uplifting terms as 
 \begin{eqnarray}
  \Delta V_i=\frac{1}{2}(\xi^{\rm FI}_i)^2g^2_{X_i}\simeq\frac{1}{2}|V_{\rm AdS}|\left(\frac{\sigma_{\tilde{i}}}{\rho_i}\right)^3
 \label{cosmo_const}
 \end{eqnarray}
 such that the value of the potential at the new minimum become equal to the observed value of the cosmological constant.  So, the anomalous FI terms can not be cancelled, and act as uplifting potential. 
And expanding the Kahler potential $K$ in components, the term linear in $V_{X_i}$ produces the FI factors $\xi^{\rm FI}_i=\frac{\partial K}{\partial V_{X_i}}\big|_{V_{X_i}=0}\,\Delta\rho_i$ as
 \begin{eqnarray}
  \xi^{\rm FI}_i=M^2_P\frac{\delta^{\rm GS}_i}{8\pi^2\sigma_{\tilde{i}}}\Delta\rho_i\,.
 \label{FI}
 \end{eqnarray}
Here the displacements $\Delta\rho_i\equiv\rho_i-\sigma_{\tilde{i}}$ in the moduli fields are induced by the uplifting terms, 
 \begin{eqnarray}
  \Delta\rho_i\simeq\frac{3M^2_P|V_{\rm AdS}|}{W^2_{TT}(\sigma_0)}\frac{\sigma_{\tilde{i}}}{\rho_i}\,,
 \end{eqnarray}
which are achieved by $\partial_{\rho_i}(V_F+\Delta V_i)=0$. Since the uplifting terms by $\Delta\rho_i$ making the dS minimum induce SUSY breaking, all particles whose mass is protected from supersymmetry become massive. 
With the choice of parameters above Eq.\,(\ref{m_TT}), the gravitino mass in Eq.\,(\ref{gravitino}) corresponds to 
 \begin{eqnarray}
  m_{3/2}\simeq560\,{\rm TeV}\,,
  \label{gravi32}
 \end{eqnarray}
impling $|\Delta W|\simeq10^{-11}M^3_P$, and which in turn means that the FI terms proportional to $|V_{\rm AdS}|/m^2_T$ are expected to be strongly suppressed.

The cosmological constant $\Lambda_c$ has the same effect as an intrinsic energy density of the vacuum $\rho_{\rm vac}=\Lambda_c M^2_P$. The dark energy density of the universe, $\Omega_\Lambda=\rho_{\rm vac}/\rho_c$, is expressed in terms of the critical density required to keep the universe spatially flat $\rho_c=3H^2_0M^2_P$ where $H_0=67.74\pm0.46\,{\rm km}\,{\rm s}^{-1}{\rm Mpc}^{-1}$ is the present Hubble expansion rate\,\cite{Ade:2015xua}.
Using the dark energy density of the universe $\Omega_\Lambda=0.6911\pm0.0062$ of Planck 2015 results\,\cite{Ade:2015xua}, 
then one finds the cosmological constant $\Lambda_c\sim7.51\times10^{-121}\,M^2_P$. From Eqs.\,(\ref{AdS}) and (\ref{cosmo_const}), one can fine-tune the value of the potential in its minimum, $V_{\rm min}$, to be equal to the observed tiny values $7.51\times10^{-121}M^4_P$,
 \begin{eqnarray}
  V_{\rm min}&=&|V_{\rm AdS}|\Big\{-1+\frac{1}{2}\left(\frac{\sigma_{\tilde{1}}}{\rho_1}\right)^3+\frac{1}{2}\left(\frac{\sigma_{\tilde{2}}}{\rho_2}\right)^3\Big\}\,.
 \label{cosmo_const1}
 \end{eqnarray}
The positive vacuum energy density resulting from a cosmological constant implies a negative pressure, and which drives an accelerated expansion of the universe, as observed.

\subsection{Moduli backreaction on inflation}
\label{inf_backRe}
Since in general the interference between the moduli and inflaton sectors generates a correction to the inflationary potential we consider the effect of string moduli backreaction on the inflation model which is linked to SUSY breaking scale\,\footnote{There are many studies\,\cite{modBack,modBack2} on the moduli backreaction effect on the inflation and its link to SUSY breaking.}. 
In small-field inflation, such as hybrid inflation, this produces a linear term in the inflaton at leading order as in Ref.\,\cite{modBack}. This is analogous to the effect of supersymmetry breaking which induces a linear term proportional to the gravitino mass. Depending on its size such a linear term can have a significant effect on inflationary
observables well fitted in CMB data, in particular, the spectral index of scalar fluctuations.

At $T_{(i)}=\bar{T}_{(i)}=\sigma_0$ due to $W(\sigma_0)=0=W_T(\sigma_0)$ one can obtain 
 \begin{eqnarray}
   V_{F}\big|_{\sigma_0}=\frac{V_{\rm inf}}{(2\sigma_0)^3}+\frac{3e^{\tilde{K}/M^2_P}}{(2\sigma_0)^3M^2_P}|W_{\rm inf}|^2\,,
 \label{Vinf_F_0}
 \end{eqnarray}
where $V_{\rm inf}$ is the inflation potential in the absence of moduli sectors
 \begin{eqnarray}
   V_{\rm inf}=e^{\tilde{K}/M^2_P}\Big\{K^{j\bar{j}}|D_{j} W_{\rm inf}|^2-\frac{3}{M^2_P}|W_{\rm inf}|^2\Big\}\,.
 \label{Vinf_F}
 \end{eqnarray}
Since all powers of $2\sigma_0$ in Eq.\,(\ref{Vinf_F_0}) can be absorbed by a redefinition of $W_{\rm inf}$ the potential is rescaled as $ V_{F}\big|_{\sigma_0}\rightarrow V_{\rm inf}+\frac{3e^{\tilde{K}/M^2_P}}{M^2_P}|W_{\rm inf}|^2$, indicating that there is no backreaction to the inflation on the moduli sector.
However, due to the effect of the inflationary large positive energy density, see Eq.\,(\ref{total energy_universe}), the minimum of the moduli are shifted by $\delta T$ and $\delta T_i$, and at this new shifted position the potential is minimized. The displacements are obtained by imposing $\partial_T V|_{\sigma_0+\delta T}=0$ and $\partial_{T_i} V|_{\sigma_0+\delta T_i}=0$, and the expression for $\delta T$ and $\delta T_i$ can be expanded in powers of $H_I/m_T$, 
 \begin{eqnarray}
  \delta T_{(i)}&\simeq&\frac{W_{\rm inf}\sqrt{3}}{2\sqrt{\sigma_0}\,m_T M^2_P}
   +\frac{1}{2(2\sigma_0)^2\,m^2_T M^2_P}\Big\{K^{j\bar{j}}D_j W_{\rm inf}\partial_{\bar{j}}\bar{W}_{\rm inf}-\frac{3}{M^2_P}|W_{\rm inf}|^2\nonumber\\
   &-&\frac{W^2_{\rm inf}}{M^2_P}\left(\frac{3}{2}+\frac{(3\sigma_0)^{3/2}W_{TTT}(\sigma_0)}{M^2_P\,m_T}\right)\Big\}+{\cal O}\left(\frac{H^3_I}{m^3_T}\right)\,.
 \label{moduli_effect}
 \end{eqnarray}
This implies that there is a supersymmetry breakdown by the inflaton sector during inflation
 \begin{eqnarray}
   D_{T_{(i)}}W\big|_{\sigma_0+\delta T_{(i)}}=\frac{1}{\sqrt{6}(2\sigma_0)^{\frac{5}{2}}\,m_T}K^{j\bar{j}}\,D_{j}W_{\rm inf}\,\partial_{\bar{j}}\bar{W}_{\rm inf}+{\cal O}\left(\frac{H^2_I}{m^2_T}\right)\,,
 \label{susy_break1}
 \end{eqnarray}
{\it i.e.}, $D_{T_{(i)}}W\big|_{\sigma_0+\delta T_{(i)}}$ are suppressed by one power of $m_T$, which vanish in the limit of infinitely heavy moduli.
 
 Since the moduli are very heavy they stabilize quickly to their minima and the inflationary potential get corrected after setting $T$ and $T_i$ to their minima as follows
 \begin{eqnarray}
   V_F\big|_{\sigma_0+\delta T_{(i)}}&=&\frac{V_{\rm inf}}{(2\sigma_0)^3}-\frac{5}{2(2\sigma_0)^5W_{TT}(\sigma_0)}\Big[W_{\rm inf}\big\{V_{\rm inf}+\frac{e^{\frac{\tilde{K}}{M^2_P}}}{5}K^{j\bar{j}}\partial_{j}W_{\rm inf}D_{\bar{j}}\bar{W}_{\rm inf}\big\}+{\rm h.c.}\Big]\nonumber\\
&+&{\cal O}\left(\frac{H^3_I}{m^3_T}\right)\,.
 \label{moduli_V}
 \end{eqnarray}
 Using $|W_{TT}(\sigma_0)|=\sqrt{\frac{2}{3}}\frac{M^2_P}{\sqrt{2\sigma_0}}m_T$, and rescaling as $V_{\rm inf}/(2\sigma_0)^3\rightarrow V_{0}(t_I)$ and $W_{\rm inf}/(2\sigma_0)^{3/2}\rightarrow W_{\rm inf}(t_I)$,  it is evident that the inflationary potential due to the moduli backreaction induces a linear term in the inflaton potential
 \begin{eqnarray}
   V_F|_{\sigma_0+\delta T_{(i)}}=V_0(t_I)\Big\{1-\frac{5\sqrt{3}}{2\sqrt{2}}\frac{1}{m_TM^2_P}(W_{\rm inf}+\text{h.c.})\Big\}
+{\cal O}\left(\frac{|\Psi_0|^3}{m^3_T}\right)
 \label{moduli_V00}
 \end{eqnarray}
Clearly, as we can see here, in the limit $m_T\rightarrow\infty$ the interference term between string moduli and inflaton sectors is disappeared.

\subsection{Scale of PQ-symmetry breakdown during inflation}
\label{inf_PQ1}
In the following, let us consider the PQ phase transition scale during inflation. Due to Eq.\,(\ref{hierarchy_vev}) during inflation we have
 \begin{eqnarray}
 v_{\Theta}(t_I)=v_{S}(t_I)=v_{T}(t_I)=0\,.
 \label{dI_v}
\end{eqnarray}
And the Kahler moduli fields we consider are stabilized during inflation and their potential has a local minimum at finite moduli fields values separated by a high barrier from the runaway direction. Since the moduli masses are much larger than the inflaton mass and accordingly will be frozen quickly during inflation without perturbing the inflaton dynamics, the height of barrier protecting metastable Minkowski ($\simeq$ dS) space are independent of the gravitino mass hence the inflationary Hubble constant is also independent of the gravitino mass\,\cite{Kallosh:2004yh}.

We consider the PQ symmetry breaking scale, $\mu_\Psi(t_I)$, during inflation.
In the global SUSY minima where $V_{\rm SUSY}=0$, all the flavon and driving fields have trivial VEVs, while the waterfall fields $\Psi$($\tilde{\Psi}$) can have non-zero VEVs. The FI $D$-terms must then be zero, {\it i.e.} $\xi^{\rm FI}_{1}=\xi^{\rm FI}_{2}=0$.  
During inflation, if $|\Psi_0|$ takes a large value the waterfall fields stay at the origin of the field space (the local minimum appears at $\langle\Psi\rangle=\langle\tilde{\Psi}\rangle=0$); and the superpotential is effectively reduced to 
 \begin{eqnarray}
 W_{\rm inf}(t_I)=-\tilde{g}_7\,\Psi_0\,\mu^2_\Psi(t_I)\,,
 \label{Winf}
\end{eqnarray}
with $\tilde{g}^2_7\equiv g^2_7/(2\sigma_0)^3$ and $\tilde{g}_7<0$, which gives a positive contribution to the inflation energy 
 \begin{eqnarray}
 V_0(t_I)=3H^2_I\,M^2_P\simeq\Big|\frac{\partial W_{\rm inf}(t_I)}{\partial\Psi_0}\Big|^2=\tilde{g}^2_7\mu^4_{\Psi}(t_I)\,,
 \label{total energy_universe}
\end{eqnarray}
and in turn drives inflation. 
Since the potential for $|\Psi_0|\gg |\Psi^c_0|\equiv\mu_{\Psi}(t_I)$ with $\langle\Psi\rangle=\langle\tilde{\Psi}\rangle=0$ is flat before the waterfall behavior occurs, inflation takes place there. And the waterfall behavior is triggered, when the inflaton $\Psi_0$ reaches the critical value $|\Psi^c_0|$. 
Once $|\Psi_0|$ rolls down from a large scale and approaches its critical value $|\Psi^c_0|$, the inflaton and waterfall fields get almost maximally mixed to form mass eigenstates:
\begin{eqnarray}
 \Psi'_0\simeq\frac{1}{\sqrt{2}}(\Psi_0\pm\tilde{\Psi})\,,\qquad\Psi'\simeq\frac{1}{\sqrt{2}}(\Psi-\Psi_{0\perp})\,,\qquad\tilde{\Psi}'\simeq-\frac{1}{\sqrt{2}}(\Psi+\Psi_{0\perp})\,,
 \label{mix_st}
\end{eqnarray}
where $\Psi_{0\perp}\simeq(\pm\Psi_0-\tilde{\Psi})/\sqrt{2}$ is orthogonal to $\Psi'_0$. And their corresponding mass eigenvalues are given by 
 \begin{eqnarray}
m_{\Psi'_0}\simeq|\tilde{g}_7|\mu_\Psi(t_I)\,,\qquad m_{\tilde{\Psi}'}\simeq|\tilde{g}_7|\mu_\Psi(t_I)\,,  \qquad m_{\Psi'}\simeq0. 
  \label{inf_ma}
 \end{eqnarray}
 
Let us schematically see this is the case.
The potential at global SUSY limit
 \begin{eqnarray}
  V^{\rm global}_{\rm inf}&=&\tilde{g}^2_7|\Psi\tilde{\Psi}-\mu^2_\Psi(t_I)|^2+\tilde{g}^2_7|\Psi_0|^2(|\Psi|^2+|\tilde{\Psi}|^2)\nonumber\\
 &=&\left(\begin{array}{cc}
 \Psi'^{\ast} & \tilde{\Psi}' \end{array}\right)\left(\begin{array}{cc}
 \tilde{g}^2_7(|\Psi_0|^2-\mu^2_\Psi(t_I)) & 0 \\ 0 & \tilde{g}^2_7(|\Psi_0|^2+\mu^2_\Psi(t_I))  \end{array}\right)
 \left(\begin{array}{c}
 \Psi' \\
 \tilde{\Psi}'^\ast \end{array}\right)+...
  \label{vglo}
 \end{eqnarray}
implies that (i) when $|\Psi_0|<\mu_{\Psi}(t_I)$, one of the mass eigenstates, $\Psi'$, becomes tachyonic: the waterfall fields fixed at $\langle\Psi\rangle=\langle\tilde{\Psi}\rangle=0$ is not stable since $\Psi(\tilde{\Psi})$ have an opposite sign of $U(1)_{X_2}$ charges.
As can be seen from Eq.\,(\ref{Kahler0}) since the Kahler moduli superfields putting the GS mechanism into practice are not separated from the SUSY breaking by the inflaton sector during inflation, by taking tachyonic SUSY breaking scalar masses $m^2_{\Psi}\sim-H^2_I$ induced dominantly by the $U(1)_{X_2}$ $D$-term, the waterfall field $\Psi'$ rolls down its true minimum from a large scale.
(ii) The other $\tilde{\Psi}'$ stays positive definite throughout the inflationary trajectory up to a critical value $|\Psi^c_0|\approx\mu_\Psi(t_I)$. (iii) After inflation the universe is dominated by both the inflaton $\Psi'_0$ and one of waterfall fields, $\tilde{\Psi}'$, while the other waterfall field $\Psi'$ gives negligible contribution to the total energy of the universe. 
(iv) After inflation and the waterfall transition mechanism has been completed $\Psi'_0$ approaches to zero and $\Psi'(\tilde{\Psi}')$ relax to the flat direction of the field space given by $\Psi'\tilde{\Psi}'=\mu^2_\Psi(t_I)$: the positive false vacuum of the inflaton field breaking the global SUSY spontaneously gets restored once inflation has been completed.
 
Now, we discuss how the inflation could be realized explicitly.
The $F$-term scalar potential, the first term in the right hand side of Eq.\,(\ref{scapot}), can be expressed as 
 \begin{eqnarray}
  V(\phi_\alpha)=e^{\tilde{K}/M^2_{\rm P}}\left\{\sum_{\alpha}K^{\alpha\bar{\alpha}}D_{\alpha} W_{\rm inf} D_{\alpha^\ast} W^\ast_{\rm inf}-3\frac{|W_{\rm inf}|^2}{M^{2}_{\rm P}}\right\}
  \label{sugraP}
 \end{eqnarray}
with $\alpha$ being the bosonic components of the superfields $\hat{\phi}_\alpha\in\{\hat{\Psi}_0$, $\hat{\Phi}^T_0$, $\hat{\Phi}^S_0$, $\hat{\Theta}_0$, $\hat{\Psi}$, $\hat{\tilde{\Psi}}$, $\hat{\Phi}_S$, $\hat{\Theta}$, $\hat{\tilde{\Theta}}$, $\hat{\Phi}_T\}$, and where the Kahler covariant derivative and Kahler metric are defined as
 \begin{eqnarray}
  D_{\alpha}W_{\rm inf}\equiv\frac{\partial W_{\rm inf}}{\partial\phi_\alpha}+M^{-2}_{\rm P}\frac{\partial K}{\partial\phi_\alpha}W_{\rm inf},\qquad K_{\alpha\bar{\beta}}\equiv\frac{\partial^2K}{\partial\phi_\alpha\partial\phi^\ast_\beta}
  \label{}
 \end{eqnarray}
and $D_{\alpha^\ast}W^\ast_{\rm inf}=(D_{\alpha}W_{\rm inf})^\ast$ with $\tilde{K}^{\alpha\bar{\beta}}\equiv (\tilde{K}_{\alpha\bar{\beta}})^{-1}$.
The lowest order ({\it i.e.} global supersymmetric) inflationary $F$-term potential $V^{\rm global}_{\rm inf}$ receives corrections for $|\phi_\alpha|\ll M_P$.
During inflation, working along the direction $|\Psi|=|\tilde{\Psi}|=0$, from Eqs.\,(\ref{NK}) and (\ref{sugraP}) a small curvature needed for the slow-roll can be represented by the inflationary potential $V_{\rm inf}$
 \begin{eqnarray}
  V_{\rm inf}&=& V^{\rm tree}_{\rm inf}+V_{\rm sugra}+\Delta V^{\rm 1-loop}_{\rm inf}\,.
  \label{V_inf}
 \end{eqnarray}
The leading order potential corrected by the interference term induced by the moduli backreaction, including soft-SUSY breaking terms associated with $\Psi_0$, can be written in Eq.\,(\ref{moduli_V00}) as
 \begin{eqnarray}
  V^{\rm tree}_{\rm inf}=V_0(t_I)\Big\{1+\frac{5\sqrt{3}}{2\sqrt{2}}\frac{\sqrt{V_0}}{m_TM^2_P}(\Psi_0+\Psi^\ast_0)\Big\}+m^2_{\Psi_0}|\Psi_0|^2-(\tilde{g}_7\,a_{\Psi_0}\mu^2_{\Psi}\Psi_0+{\rm h.c.})\,,
  \label{V_0}
 \end{eqnarray}
where $V_0(t_I)$ is the rescaled vacuum energy during inflation, see Eq.\,(\ref{moduli_V00}), and $a_{\Psi_0}$ is the soft-SUSY breaking mass parameter of order $\sim m_{3/2}$.  In Eq.\,(\ref{V_0}) we only have included the tadpole term since all other soft-SUSY breaking terms are negligible during inflation.
Substituting $K_{\rm inf}$ and $W_{\rm inf}$ in Eq.\,(\ref{NK}) into $V^{\rm inf}_F$ in Eq.\,(\ref{Vinf_F}), and minimizing with respect to $\Psi$ and $\tilde{\Psi}$ for $|\Psi_0|>\mu_\Psi(t_I)$ gives
 \begin{eqnarray}
  V^{\rm inf}_F&=& \tilde{g}^2_7\mu^4_{\Psi}(t_I)\left\{1-k_{s}\frac{|\Psi_0|^2}{M^2_{P}}+\gamma_{s}\frac{|\Psi_0|^4}{2M^4_{P}}+{\cal O}\Big(\frac{|\Psi_0|^6}{M^6_P}\Big)\right\}\,,
  \label{V_sugra0}
 \end{eqnarray}
where $\gamma_{s}\equiv1-7k_{s}/2-3k_{3}$. Such a supergravity induced mass squared is expected to have the same form as the $\Psi_0$ mass squared, namely $\tilde{g}^2_7\mu^4_\Psi(t_I)/M^2_P=V_0(t_I)/M^2_P$ which is the order of the Hubble constant squared $H^2_I=V_0(t_I)/3M^2_P$.
Then the SUGRA contribution $V_{\rm sugra}$ to $V_{\rm inf}$ leads to 
 \begin{eqnarray}
  V_{\rm sugra}&=& -c^2_H H^2_I|\Psi_0|^2+V_0\gamma_{s}\frac{|\Psi_0|^4}{2M^4_{P}}+{\cal O}\Big(\frac{|\Psi_0|^6}{M^6_P}\Big)\,.
  \label{V_sugra}
 \end{eqnarray}
The inflaton $\Psi_0$ also receives 1-loop radiative correction in the potential\,\cite{Coleman:1973jx}  due to the mismatch between masses of the scalar and fermion components of $\Psi(\tilde{\Psi})$, which are non-vanishing since SUSY is broken by $\partial W_{\rm inf}/\partial\Psi_0\neq0$. The corresponding 1-loop correction to the scalar potential is analytically calculated as
 \begin{eqnarray}
  &&\Delta V_{\rm 1-loop}=\sum_i(-1)^f\frac{m^4_i}{64\pi^2}\ln\frac{m^2_i}{Q^2}=\frac{\tilde{g}^4_7\,\mu^4_{\Psi}(t_I)}{8\pi^2}F(x)
  \label{1-loo}
 \end{eqnarray}
 where $F(x)=\frac{1}{4}\big\{(x^2+1)\ln\frac{x^4-1}{x_4}+2x^2\ln\frac{x^2+1}{x^2-1}+2\ln\frac{g^2_7\mu^2_\Psi\,x^2}{Q^2}-3\big\}$ and the sum is taken over the field degrees of freedom and $f=0$ for scalar and $f=1$ for fermion. Here the $Q$ is a renormalizable  scale, $x$ is defined as $x\equiv|\Psi_0|/\mu_{\Psi}(t_I)=\varphi/(\sqrt{2}\,\mu_\Psi(t_I))$ where $\varphi$ is the normalized real scalar field. In the limit $x\gg1$, i.e. $\varphi\gg\sqrt{2}\,\mu_\Psi(t_I)$, this is approximated as
  \begin{eqnarray}
  \Delta V_{\rm 1-loop}&\simeq&\frac{\tilde{g}^4_7\,\mu^4_\Psi(t_I)}{16\pi^2}\ln\frac{\tilde{g}^2_7\,\varphi^2}{2Q^2}\,.
  \label{}
 \end{eqnarray}

If we let the inflaton field $\Psi_0\equiv\varphi\,e^{i\theta}/\sqrt{2}$, and during the inflation period, taking into account the radiative correction, supergravity effects, soft-SUSY breaking terms, and moduli backerction effects, the inflationary potential is of the following form
 \begin{eqnarray}
  V_{\rm inf}(\varphi)&=&V_0(t_I)\Big\{1+\frac{5\sqrt{3}}{2}\frac{\sqrt{V_0}}{m_TM^2_P}\varphi\cos\theta+\gamma_s\frac{\varphi^4}{8M^4_P}+\frac{\tilde{g}^2_7}{8\pi^2}F(x)\Big\}\nonumber\\
  &+&\tilde{g}_7\,\alpha_s\,m_{3/2}\,\mu^2_\Psi\varphi\cos\theta+\frac{\varphi^2}{2}\left(m^2_{\Psi_0}-k_s\frac{V_0}{M^2_P}\right)\,,
  \label{pot_inf}
 \end{eqnarray}
 where $\alpha_s\,m_{3/2}=-\sqrt{2}\,a_{\Psi_0}$.
The moduli-induced slope partially cancels the slope of the Coleman-Weinberg potential, which flattens the inflationary trajectory and reduces the distance in field space corresponding to the $N_e\sim50$ $e$-folds of inflation. And the inflaton mass $m_{\Psi_0}$ is assumed for $k_s=1$ as 
 \begin{eqnarray}
  m_{\Psi_0}=|\tilde{g}_7|\frac{\mu^2_\Psi(t_I)}{M_P};
  \label{inflaton_mass}
 \end{eqnarray}
since the inflaton acquires a mass of order the Hubble constant, $m_{\Psi_0}=H_I\sqrt{3}$, agreement of theory's prediction for spectral index $n_s$ with observation strongly suggests the presence of a negative Hubble-induced mass-term, and the $k_s$ parameter term vanishes identically. This inflaton mass ($\gg{m}_{3/2}$) can directly be obtained from Eqs.\,(\ref{NK0}) and (\ref{NK}) as
\begin{eqnarray}
 m_{\Psi_0}=\big|M^4_P\langle{e}^{G}\nabla_{\Psi_0}G_{\Psi_0}\rangle\big|^{\frac{1}{2}}=\sqrt{3}H_I\,,
 \label{infl_ma}
\end{eqnarray}
where $\nabla_k{G}_\alpha=\partial_k{G}_\alpha-\Gamma^j_{k\alpha}G_j$ with the Christoffel symbol $\Gamma^j_{k\alpha}=G^{j\ell^\ast}G_{k\alpha\ell^\ast}$\,\cite{wbg}, and $\nabla_{\Psi_0}G_{\Psi_0}\simeq-(W_{\Psi_0}/W)^2$ is used. This inflaton mass is in agreement with the above prediction in Eq.\,(\ref{inflaton_mass}).

Inflation stops at $|\Psi^c_0|\simeq\mu_\Psi(t_I)$, where the mass of $\Psi$ becomes negative and the field acquires a non-vanishing expectation value.
In order to develop the VEV of the waterfall field $\Psi$, we destabilize the waterfall field $\Psi$ by taking tachyonic Hubble induced masses of the PQ-breaking waterfall field, {\it i.e.}, $m^2_{\Psi}\sim-H^2_I<0$. Then, the VEV of the waterfall field could be determined by considering both the SUSY breaking effect and a supersymmetric next leading order term. The next leading Planck-suppressed operator invariant under $A_4\times U(1)_X$ is given by
 \begin{eqnarray}
 \Delta W_{v}&\simeq& \frac{\hat{\alpha}}{M^2_{P}}\Psi_0\Psi^2\tilde{\Psi}^2\,,
 \label{NPpotential1}
 \end{eqnarray}
where we set the VEVs of all other matter fields to zero except the waterfall field and neglected their corresponding trivial operators. Note that the constant $\hat{\alpha}={\cal O}(\alpha/8\pi)$ with a constant $\alpha$ being of order unity.
Since the soft SUSY-breaking terms are already present at the scale relevant to inflation dynamics, the scalar potential for the waterfall field $\Psi$ at leading order reads
 \begin{eqnarray}
 V_{\Psi}(t_I)
 &\simeq&\frac{1}{2}D^2_{X_2}+\hat{\alpha}_{\Psi}\,\tilde{m}^2_{\Psi}|\Psi|^2+\hat{\alpha}_{\tilde{\Psi}}\,\tilde{m}^2_{\tilde{\Psi}}|\tilde{\Psi}|^2+|\hat{\alpha}|^2\,\frac{|\Psi|^{4}|\tilde{\Psi}|^{4}}{M^4_{P}}+...\,,
 \label{HI_poten}
\end{eqnarray}
where $|\hat{\alpha}_{\Psi}\,\tilde{m}^2_{\Psi}|,|\hat{\alpha}_{\tilde{\Psi}}\,\tilde{m}^2_{\tilde{\Psi}}|$$\ll |D_{X_2}(t_I)|$ with $|\hat{\alpha}_{\Psi,\tilde{\Psi}}|\ll1$ are taken.
Here $\tilde{m}_{\Psi,\tilde{\Psi}}\simeq|\Psi^c_0|\sim{\cal O}(|F^{\Psi_0}|/M_P)$ with $F^{\Psi_0}=K^{\Psi_0\bar{\Psi}_0}\,D_{\Psi_0}W_{\rm inf}\simeq\sqrt{3}\,H_I\,M_P$ represents the Hubble induced soft scalar masses generated by the $F$-term SUSY breaking, during inflation.
If the tachyonic SUSY breaking scalar masses are dominantly induced by the $U(1)_{X_2}$ $D$-term, $D_{X_2}(t_I)\sim{\cal O}(H^2_I)$, compared to the Hubble induced soft masses generated by the $F$-term SUSY breaking, the soft SUSY breaking mass of $\Psi$ during inflation are approximated by 
 \begin{eqnarray}
  m^2_{\Psi}(t_I)=\hat{\alpha}_{\Psi}\,\tilde{m}^2_{\Psi}+D_{X_2}(t_I)\simeq-\hat{\beta}_{\Psi}\,H^2_I\,,\quad\text{with}~\hat{\beta}_{\Psi}>0\,.
\end{eqnarray}
Then the scalar potential in Eq.\,(\ref{HI_poten}) for the waterfall field $\Psi$ is good approximated as
 \begin{eqnarray}
 V_{\Psi}(t_I)
 &\simeq&-\hat{\beta}_{\Psi}\,H^2_{I}|\Psi|^2+|\hat{\alpha}|^2\,\frac{|\Psi|^{4}|\tilde{\Psi}|^{4}}{M^4_{P}}\,.
 \label{}
\end{eqnarray}
Here the constant $\hat{\beta}_{\Psi}$ are of order unity, while $\hat{\alpha}=\alpha/(8\pi)$ with $\alpha$ being of order unity.
We find the minimum as
 \begin{eqnarray}
  v_{\Psi}(t_I)=\sqrt{\frac{2\hat{\beta}_{\Psi}}{|\hat{\alpha}|^2}}\,H_I\left(\frac{M_P}{v_{\tilde{\Psi}}}\right)^2\,,
 \label{inf_f2}
\end{eqnarray}
leading to $M_P\gg\mu_\Psi(t_I)\gg H_I$ and the PQ breaking scales during inflation
\begin{eqnarray}
  \mu^2_{\Psi}(t_I)&\equiv&\frac{v_{\Psi}(t_I)\,v_{\tilde{\Psi}}(t_I)}{2}=\sqrt{\frac{\hat{\beta}_{\Psi}}{2|\hat{\alpha}|^2}}\,\left(\frac{H_{I}}{v_{\tilde{\Psi}}(t_I)}\,M^2_{P}\right)\,.
 \label{PQ_scale2}
\end{eqnarray}
In supersymmetric theories based on SUGRA, since SUSY breaking is transmitted by gravity, all scalar fields acquire an effective mass of the order of the expansion rate during inflation. So, we expect that the inflaton acquires a mass of order the Hubble constant, and which  in turn indicates that the soft SUSY breaking mass (the inflaton mass $m_{\Psi_0}$) during inflation strongly depends on the scale of waterfall (or PQ) fields by the above Eq.\,(\ref{PQ_scale2}); for example, for $\mu_{\Psi}(t_I)\sim10^{16}$ GeV one obtains 
\begin{eqnarray}
  H_{I}\sim2\times10^{10}\,\text{GeV}
\end{eqnarray}
for $\hat{\beta_i}\sim1$ and $\hat{\alpha}\sim1/(8\pi)$, see TABLE\,\ref{inf_result}. 

After the inflation ends, for simplicity, we treat the mixed mass eigenstates in Eq.\,(\ref{mix_st}) as the single field eigenstates, 
\begin{eqnarray}
 \Psi'_0\rightarrow\Psi_0\,,\qquad \Psi'\rightarrow\Psi\,,\qquad \tilde{\Psi}'\rightarrow\tilde{\Psi}\,.
 \label{reexp}
\end{eqnarray}
Then, we express the superpotential (\ref{NK0}) relevantly 
\begin{eqnarray}
 W\supset W(z)+\tilde{g}_7\Psi_0(\tilde{\Psi}\Psi-\mu^2_\Psi)
 \label{supWz}
\end{eqnarray}
where $W(z)$ is introduced to determine SUSY breaking scale, see Sec.\,\ref{susy_break}, and $\tilde{g}^2_7=g^2_7/(2\sigma_0)^{3}$ corrected by the string moduli backreaction. Then the scalar potential in Eq.\,(\ref{scapot}) is extremized in the true vacuum if $\langle\partial_iV\rangle=0$, and the resulting cosmological constant should vanish if $\langle V\rangle=0$. Together with, these conditions are satisfied if
\begin{eqnarray}
 \langle{G}^\alpha{G}_\alpha\rangle=3\,,\qquad\qquad\langle{G}^\alpha\nabla_k{G}_\alpha+G_k\rangle=0\,.
\label{pom}
\end{eqnarray}
 Then the condition of the potential minimum read
\begin{eqnarray}
&& \langle M^2_P\{{G}_{\Psi_0\Psi_0}{G}_{\bar{\Psi}_0}+{G}_{\Psi\Psi_0}{G}_{\bar{\Psi}}+{G}_{\tilde{\Psi}\Psi_0}{G}_{\bar{\tilde{\Psi}}}+{G}_{z\Psi_0}{G}_{\bar{z}}\}+G_{\Psi_0}\rangle=0\,,\label{mcon1}\\
&& \langle M^2_P\{{G}_{\Psi\Psi}{G}_{\bar{\Psi}}+{G}_{\Psi_0\Psi}{G}_{\bar{\Psi_0}}+{G}_{\tilde{\Psi}\Psi}{G}_{\bar{\tilde{\Psi}}}+{G}_{z\Psi}{G}_{\bar{z}}\}+G_{\Psi}\rangle=0\,,\label{mcon2}
\end{eqnarray}
and the minimization condition for $\tilde{\Psi}$ is the same as for $\Psi$. 
The inflaton mass ($\gg{m}_{3/2}$), after the inflation, is given by
\begin{eqnarray}
 m_{\Psi_0}\simeq\big|M^4_P\langle{e}^{G}\nabla_{\Psi_0}G_{\tilde{\Psi}}\nabla_{\Psi_0}G^{\tilde{\Psi}}\rangle\big|^{\frac{1}{2}}\simeq|\tilde{g}_7|\mu_\Psi(t_I)\,,
 \label{infl_ma1}
\end{eqnarray}
where $\nabla_{\Psi_0}G_{\tilde{\Psi}}\simeq W_{\tilde{\Psi}\Psi_0}/W$ is used, which is almost equal to the mass of waterfall field $\tilde{\Psi}$. This inflaton mass is in agreement with Eq.\,(\ref{inf_ma}).
Since the $z$ field is responsible for the SUSY breaking, one obtains $|G_z|\simeq\sqrt{3}/M_P$, and in turn the gravitino mass $m_{3/2}\equiv\langle M_P\,{e}^{G/2}\rangle\simeq|W|/M^2_P\simeq|W_z|/\sqrt{3}M_P$. Assuming $|G_\Psi|\simeq|G_{\tilde{\Psi}}|\lesssim|\Psi|/M^2_P$, one obtains $G_\Psi\simeq{W}_{\Psi}/W$, leading to $W_\Psi/W\simeq\Psi/M^2_P$ and $W_{\tilde{\Psi}}/W\simeq\tilde{\Psi}/M^2_P$. Using $W_\Psi=\tilde{g}_7\Psi_0\tilde{\Psi}$ in Eq.\,(\ref{supWz}) we obtain
\begin{eqnarray}
 \langle\Psi_0\rangle\simeq\frac{m_{3/2}}{|\tilde{g}_7|}\,.
 \label{ve0}
\end{eqnarray}

\subsection{Cosmological observables}
\label{inf_PQ2}
The inflaton as a source of inflation is displaced from its minimum and whose slow-roll dynamics leads to an accelerated expansion of the early universe.
During inflation the universe experiences an approximately dS phase with the Hubble parameter $H_I$.
Quantum fluctuations during this phase can lead to observable signatures in CMB radiation temperature fluctuation, as the form of density perturbation, in several ways~\cite{Fox:2004kb}, when the quantum fluctuations are crossing back inside the Hubble radius long after inflation has been completed. When interpreted in this way, inflation provides a causal mechanism to explain the observed nearly-scale invariant CMB spectrum.
(i) Quantum fluctuations of the inflaton field during inflation give rise to fluctuations in the scalar curvature and lead to the adiabatic fluctuations\,\footnote{These correspond to fluctuations in the total energy density, $\delta\rho\neq0$, with no fluctuation in the local equation of state, $\delta(n_i/s)=0$. On the other hand, isocurvature perturbations correspond to fluctuations in the local equation of state of some species, $\delta(n_i/s)\neq0$, with no fluctuation in the total energy density, $\delta\rho=0$\,\cite{Fox:2004kb}.} that have grown into our cosmologically observed large-scale structure much bigger than the Hubble radius and then eventually got frozen.
Adiabatic density perturbations seeded by the quantum fluctuations of the inflaton have a nearly scale-invariant spectrum, $\Delta^2_{\cal R}(k_0)$,  which is a cosmological observable of the curvature perturbations. The power spectrum of the curvature perturbations, $\Delta^2_{\cal R}(k_0)$, reads in the Planck 2015 result at $68\%$ CL (for the base $\Lambda$CDM model)\,\cite{Ade:2015xua} 
  \begin{eqnarray}
  \Delta^2_{\cal R}(k_0)=(2.141^{+0.050}_{-0.049})\times10^{-9}\,,
  \label{cur_per}
 \end{eqnarray}
at the pivot scale $k_0=0.002\,{\rm Mpc}^{-1}$ (wave number), which is compatible with the one suggested for the COBE normalization\,\cite{Bunn:1996da}.
(ii) Fluctuations of the metric lead to tensor-B mode fluctuations in the CMB radiation. Primordial gravitational waves are generated with a nearly scale-invariant spectrum, $\Delta^2_{h}(k_0)$, which reads in the Planck 2015 result\,\cite{Ade:2015xua}  $\Delta^2_{h}(k_0)<1.97\times10^{-10}$. 
(iii) Quantum fluctuations are imprinted into every massless scalar field in dS space during inflation, with an approximately scale-invariant spectrum, $\langle|\delta\phi(k)|^2\rangle=(H_I/2\pi)^2/(k^3/2\pi^2)$
for a canonically normalized scalar field $\phi$, which is essentially a thermal spectrum at Gibbons-Hawking temperature $T_{\rm GH}=H_I/2\pi$. 
The other important cosmological observables imprinted in the CMB spectrum are followings: the BAU (which will be discussed in Sec.\,\ref{AD}), the fractions of relic abundance $\Omega_{\rm DM}$ (see Ref.\,\cite{Ahn:2016hbn})  and dark energy $\Omega_{\Lambda}$ (see Sec.\,\ref{susy_break}).

The slow-roll condition\,\cite{slow_roll} is well satisfied up to the critical point $\varphi^c=\sqrt{2}\,\mu_\Psi(t_I)$, beyond which the waterfall mechanism takes place. Here the slow-roll parameters, $\epsilon$ and $\eta$, are approximately derived from Eq.\,(\ref{pot_inf}) as
  \begin{eqnarray}
   \epsilon&\equiv&\frac{M^2_P}{2}\left(\frac{V_{\varphi}}{V}\right)^2\label{slow_curvature02}\\
   &\simeq&\frac{1}{2}\Big(\frac{\tilde{g}^2_7}{8\pi^2}\frac{M_P}{\varphi}\Big)^2\Big\{1+\frac{5\sqrt{3}}{2}\frac{8\pi^2}{|\tilde{g}_7|}\frac{\mu_\Psi}{m_T}\frac{\mu_\Psi}{M_P}\frac{\varphi}{M_P}\cos\theta\Big(1-\frac{\alpha_s\,m_{3/2}\varphi\cos\theta}{\tilde{g}_7\mu^2_\Psi}\Big)\nonumber\\
   &&\qquad\qquad\qquad\quad+\frac{8\pi^2\alpha_s}{\tilde{g}^3_7}\Big(\frac{m_{3/2}}{\mu_\Psi}\Big)\Big(\frac{\varphi}{\mu_\Psi}\Big)\cos\theta\Big\}^2\ll1\,,
     \label{slow_curvature1}\\
    \eta&\equiv& M^2_P\frac{V_{\varphi\varphi}}{V}\label{slow_curvature01}\\
    &\simeq&\frac{\tilde{g}^2_7}{8\pi^2}\Big(\frac{M_P}{\varphi}\Big)^2\Big\{\frac{3\gamma_s}{2}\frac{8\pi^2}{\tilde{g}^2_7}\Big(\frac{\varphi}{M_P}\Big)^2-1\Big\}\Big(1-\frac{\alpha_s\,m_{3/2}\varphi\cos\theta}{\tilde{g}_7\mu^2_\Psi}\Big)\,,~|\eta|\ll1\,,
  \label{slow_curvature}
 \end{eqnarray}
where $V_{\varphi}$ denotes a derivative with respect to the inflaton field $\varphi=\sqrt{2}\,{\rm Re}\Psi_0$, and $M_P\gg|\Psi_0|\gg|\Psi^c_0|$ (or $M_P\gg|\varphi|\gg|\varphi^c|$) is assumed. Recalling that $\tilde{g}^2_7=g^2_7/(2\sigma_0)^3$. The above equations clearly show that the curvature of the inflationary potential is dominantly affected by the moduli backreaction in Eq.\,(\ref{moduli_V00}), the 1-loop radiative correction in Eq.\,(\ref{1-loo}), and soft-SUSY breaking term in Eq.\,(\ref{V_0}). In the slow-roll approximation, the number of $e$-foldings after a comoving scale $l$ has crossed the horizon is given by the inflationary potential through
  \begin{eqnarray}
  N(\varphi)&=&\int^{t_l}_{t(\varphi^c)}H_Idt=\frac{1}{M^2_P}\int^{\varphi_l}_{\varphi^c}\frac{V(\varphi)}{V_\varphi(\varphi)}d\varphi\,,
  \label{e-foldings}
 \end{eqnarray}
where $\varphi_l$ is the value of the field at the comoving scale $l$, and $\varphi^c$ is the one at the end of inflation.
 The field value $\varphi^c$ is determined from the condition ${\rm Max}\{\epsilon(\varphi^c),|\eta(\varphi^c)|\}=1$\,\cite{Liddle:1992wi}.
The power spectrum $\Delta^2_{\cal R}(k_0)$ sensitively depends on the theoretical parameters of the inflationary potential,
  \begin{eqnarray}
  \Delta^2_{\cal R}(k_0)&\simeq&\frac{1}{12\pi^2\,M^6_P}\frac{V^3(\varphi_l)}{|V_\varphi(\varphi_l)|^2}
  \label{power_spec}
 \end{eqnarray}
where the potential $V(\varphi_l)$ and its derivative $V_\varphi(\varphi_l)$ are evaluated at the epoch of horizon exit for the comoving scale $k_0$. It should be compared with the Planck 2015 result Eq.\,(\ref{cur_per}).
With the definition of the number of $e$-folds after a comoving scale $k_0$ leaves the horizon, we can obtain the corresponding inflaton value $\varphi_l/M_P$ from Eq.\,(\ref{e-foldings}).
 And the number of $e$-folds $N_e$ corresponding to the comoving scale $k_0$ is around 50 depending on the energy scales $H_I$ and $T_{\rm reh}$
  \begin{eqnarray}
  N_e=49.1+\ln\left(\frac{0.002\,{\rm Mpc}^{-1}}{k_0}\right)+\frac{1}{3}\ln\left(\frac{T_{\rm reh}}{10^{4}\,{\rm GeV}}\right)+\frac{1}{3}\ln\left(\frac{H_I}{10^{10}\,{\rm GeV}}\right)
  \label{ef}
 \end{eqnarray}
where $T_{\rm reh}$ represents the maximal temperature of the last radiation dominated era, so-called the reheating temperature. 
 The tensor and scalar modes have spectrum $A_t=2H^2_I/(\pi^2M^2_P)$ and $A_s\equiv\Delta^2_{\cal R}(k_0)$\,\cite{Ade:2015xua}, respectively. In the supergravity $F$-term inflation we consider, the tensor-to-scalar ratio $r=A_t/A_s\simeq16\,\epsilon(\varphi_l)$ is much lower than the Planck 2015 bound ($r_{0.002}<0.09$), {\it i.e.} well bellow $10^{-2}$, and the running of the spectral index $dn_s/d\ln\tilde{g}_7$ is always smaller than $10^{-3}$ and so unobservable. 
And the scalar spectral index $n_s$ is approximated as 
  \begin{eqnarray}
  n_s\simeq1-6\,\epsilon(\varphi_l)+2\,\eta(\varphi_l)\simeq2\,\eta(\varphi_l)\,.
  \label{sindex}
 \end{eqnarray}
We can compare this quantity with the results of the Planck 2015 observation\,\cite{Ade:2015xua}
  \begin{eqnarray}
  n_s=0.967\pm0.004\,.
  \label{nsP}
 \end{eqnarray}
\begin{figure}[t]
\begin{minipage}[h]{7.5cm}
\epsfig{figure=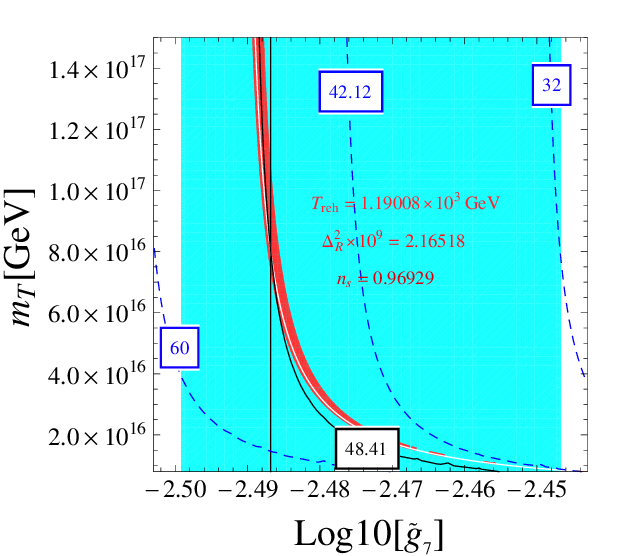,width=7.8cm,angle=0}
\end{minipage}
\hspace*{1.0cm}
\begin{minipage}[h]{7.5cm}
\epsfig{figure=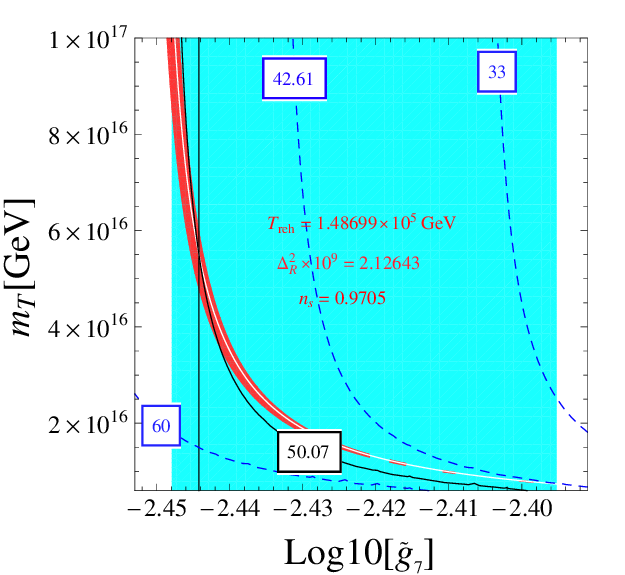,width=7.8cm,angle=0}
\end{minipage}\\
\begin{minipage}[h]{7.5cm}
\epsfig{figure=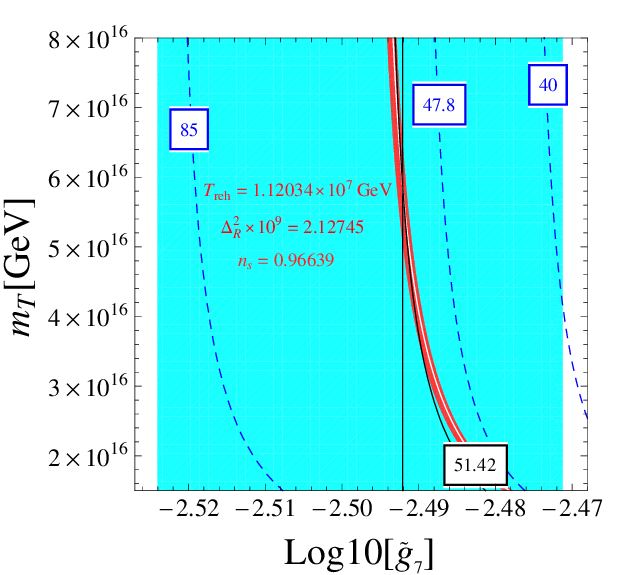,width=7.8cm,angle=0}
\end{minipage}
\hspace*{1.0cm}
\begin{minipage}[h]{7.5cm}
\epsfig{figure=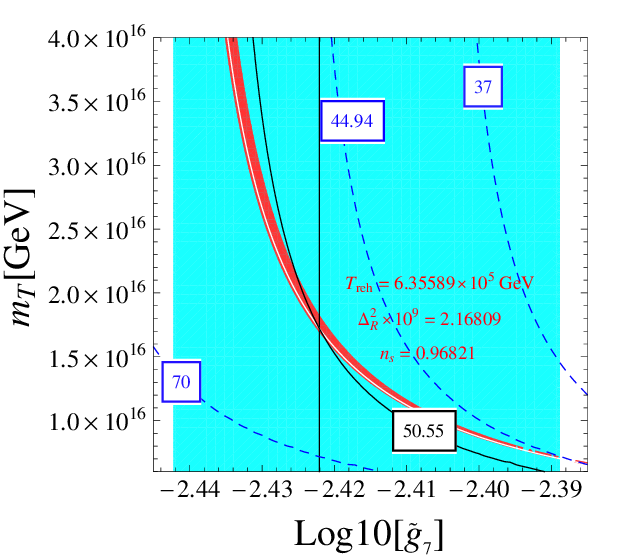,width=7.8cm,angle=0}
\end{minipage}
\caption{\label{Fig1} Contour-plot for $N_{e}$ as a function of $m_T $ and $|\tilde{g}_{7}|$ with the given values of $\alpha_s$, $\gamma_s$, $\varphi_l$, and $\varphi^c$ in TABLE-\ref{inf_result}, where each red-band curves and cyan-vertical bands stands for the allowed regions of the constraints $\Delta^2_{\cal R}(k_0)$ and $n_s$ in Eqs.(\ref{cur_per}) and (\ref{nsP}), respectively. Each intersection point among white curve ($\Delta^2_{\cal R}(k_0)$), black-solid curve ($N_e$), and black-vertical line ($n_s$) corresponds to each input values with high accuracies in TABLE-\ref{inf_result}.}
\end{figure}

In order for the power spectrum of the curvature perturbation in Eq.\,(\ref{power_spec}) and the spectral index in Eq.\,(\ref{sindex}) with Eqs.\,(\ref{slow_curvature02}) and (\ref{slow_curvature01})
 to be well fitted with the Planck 2015 observation, the five independent parameters $m_T$, $\mu_\Psi(t_I)$, $\gamma_s$, $\alpha_s$, and $|\tilde{g}_7|$ in Eq.\,(\ref{pot_inf}) are needed and those parameters with the conditions Eqs.\,(\ref{cur_per}) and (\ref{nsP}) have predictions, $m_T={\cal O}(10^{16-17})$$\gg\mu_\Psi(t_I)=\varphi^c/\sqrt{2}={\cal O}(10^{15})$ GeV, $\gamma_s={\cal O}(1-10)$, $|\alpha_s|={\cal O}(1)$, and $|\tilde{g}_7|={\cal O}(1)\times10^{-3}$ as in TABLE\,\ref{inf_result}, where we have set $\cos\theta=-1$ and $m_{3/2}=560$ TeV (see Eq.\,(\ref{gravi32})).
\begin{table}[h]
\caption{\label{inf_result} Five independent input parameters $m_T$, $\gamma_s$, $\alpha_s$, $|\tilde{g}_7|$, and $\mu_{\Psi}(t_I)=\varphi^c/\sqrt{2}$ in the inflationary potential of Eq.\,(\ref{pot_inf}) provide predictions on $N_e$ and $T_{\rm reh}/{\rm GeV}$ with the constraints $\Delta_{\cal R}(k_0)/10^{-9}$ in Eq.\,(\ref{cur_per}) and $n_s$ in Eq.\,(\ref{nsP}), where $\cos\theta=-1$ and $m_{3/2}=560$ TeV in Eq.\,(\ref{gravi32}) are taken.}
\begin{ruledtabular}
\begin{tabular}{cccc||ccc||cccc}
$\frac{m_T}{10^{16}\,\rm GeV}$ &$\alpha_s$ &$\gamma_s$&$\frac{|\tilde{g}_7|}{10^{-3}}$& $\frac{H_I}{10^{10}\rm GeV}$  & $\frac{\varphi_l}{10^{15}\rm GeV}$ & $\frac{\varphi^c}{10^{15}\rm GeV}$ &$n_s$&$\frac{\Delta_{\cal R}(k_0)}{10^{-9}}$&$N_{e}$ & $\frac{T_{\rm reh}}{\rm GeV}$ \\
\hline
$5.87238$ &$0.85492$&$1.76989$& $ 3.22039$ & $0.90076$ & $8.28466$ & $4.85231$ & $0.96639$ & $2.12745$ & $51.42363$ & $1.21034\times10^{7}$\\
$1.72083$ &$1.07814$&$-6.75512$& $ 3.78371$ & $1.26228$ & $9.78007$ & $5.29929$ & $0.96821$ & $2.16809$ & $50.55386$ & $6.35558\times10^{5}$\\
$5.51975$ &$1.06936$&$2.63451$& $ 3.59549 $ & $1.26311$ & $9.69460$ & $5.43800$ & $0.97050$ & $2.12643$ & $50.06988$ & $1.48699\times10^{5}$\\
$8.04311$ &$0.90832$&$5.88591$& $ 3.25965 $ & $1.09422$ & $8.84520$ & $5.31575$ & $0.96929$ & $2.16518$ & $48.41274$ & $1.19008\times10^{3}$\\
\end{tabular}
\end{ruledtabular}
\end{table}
This table shows that the cosmological observables can be well fitted where both the moduli stabilized at a scale close to $\Lambda_{\rm GUT}$ and the PQ symmetry breaking scale induced at $\mu_{\Psi}(t_I)\simeq0.3\times10^{16}$ GeV$<m_T$. 
Fig.\,\ref{Fig1} shows the behavior of the number of $e$-folds $N_e$ in Eq.\,(\ref{ef}) in terms of the five independent parameters of the inflationary potential in Eq.\,(\ref{pot_inf}) $m_T$, $\mu_\Psi(t_I)$, $\alpha_s$, $\gamma_s$, and $|\tilde{g}_7|$, where each red-band curves and cyan-vertical bands stands for the allowed regions of the constraints $\Delta^2_{\cal R}(k_0)$ and $n_s$ in Eqs.(\ref{cur_per}) and (\ref{nsP}), respectively. Each of the contour plot in clockwise corresponds to the value of TABLE-\ref{inf_result} in sequence from top to bottom. In the plots showing contour lines for $N_e$ in terms of the parameter set $\{m_T, |\tilde{g}_7|\}$ with the given input values of the parameter set $\{\alpha_s, \gamma_s, \mu_\Psi\}$ in TABLE-\ref{inf_result}, each of the region of red-band curve overlapped by the cyan-vertical band represents each of the allowed region by the constraints $\Delta^2_{\cal R}(k_0)$ and $n_s$ in Eqs.(\ref{cur_per}) and (\ref{nsP}), leading to large uncertainties of reheating temperature $T_{\rm reh}$ corresponding to the allowed range of $N_e$: $42.12\lesssim N_e\lesssim 48.79$ (left-upper panel), $42.61\lesssim N_e\lesssim 51.14$ (right-upper panel), $44.94\lesssim N_e\lesssim53.84$ (right-lower panel), and $47.80\lesssim N_e\lesssim52.33$ (left-lower panel) with an assumption of $m_T\leq10^{17}$ GeV. 

In the plots, especially, each intersection point among white curve ($\Delta^2_{\cal R}(k_0)$), red-solid curve ($N_e$), and red-vertical line ($n_s$) corresponds to each input values $m_T$, $|\tilde{g}_7|$, $\alpha_s$, $\gamma_s$, and $\mu_\Psi$ with such high accuracies in TABLE-\ref{inf_result}. For the given values of reheating temperature and parameter set $\{\mu_\Psi, \alpha_s, \gamma_s\}$ in TABLE-\ref{inf_result} we obtain theoretical uncertainties of $\Delta_{\cal R}(k_0)$ and $n_s$ corresponding to the theoretical uncertainties of the parameter set $\{m_T, |\tilde{g}_7|\}$: 
\begin{eqnarray}
&&\Delta_{\cal R}(k_0)/10^{-9}=2.16518^{+0.02582}_{-0.07318}\,,\qquad n_s=0.96929^{+0.00009}_{-0.00017}\,,\quad\text{for left-upper panel}\,,\nonumber\\
&&\Delta_{\cal R}(k_0)/10^{-9}=2.12643^{+0.06457}_{-0.03443}\,,\qquad n_s=0.97050^{+0.00022}_{-0.00040}\,,\quad\text{for right-upper panel}\,,\nonumber\\
&&\Delta_{\cal R}(k_0)/10^{-9}=2.16809^{+0.02291}_{-0.07609}\,,\qquad n_s=0.96821^{+0.00049}_{-0.00014}\,,\quad\text{for right-lower panel}\,,\nonumber\\
&&\Delta_{\cal R}(k_0)/10^{-9}=2.12745^{+0.06355}_{-0.03545}\,,\qquad n_s=0.96639^{+0.00024}_{-0.00041}\,,\quad\text{for left-lower panel}\,.
\label{hiac}
\end{eqnarray}
where an assumption of $m_T\leq10^{17}$ GeV is considered for the case of left-upper panel.
 Note that the high accuracies in Eq.\,(\ref{hiac}) are due to the fact that  the slow-roll parameter $\eta$ given in Eq.\,(\ref{slow_curvature}) governing the spectral index $n_s$ is very sensitive to values of the parameter $|\tilde{g}_7|$.
As shown in TABLE-\ref{inf_result}, the number of $e$-foldings in Eq.\,(\ref{ef}) depends on the amount of reheating temperature, which in turn depends on the decay rate of the inflaton $\Psi_0$ and waterfall field $\tilde{\Psi}$ into relativistic particles. In the following section we will see how the amount of reheating, $T_{\rm reh}$, could be strongly correlated with both baryogenesis via leptogenesis and the yield of gravitinos.

\section{Leptogenesis}
 \label{AD}
Let us discuss on how the matter-antimatter asymmetry of the universe could be realized in the context of the present model. In order to account for a successful leptogenesis, we introduce the AD mechanism for baryogenesis\,\cite{Affleck:1984fy} and its subsequent leptonic version so-called AD leptogenesis\,\cite{Murayama:1993em}.
 In the global SUSY limit, {\it i.e.} $M_P\rightarrow\infty$, as well as in the energy scale where $A_4\times U(1)_X$ is broken (see Ref.\,\cite{Ahn:2016hbn}), some combinations of scalar fields do not enter the potential, composing flat directions of the scalar potential. So, taking the flat directions $H_u=L_i=\zeta_i/\sqrt{2}$ (a generation index $i=1,2,3$), then the AD flat directions for leptogenesis\,\cite{Murayama:1993em} are $\zeta_i=(2\widetilde{L_i}H_u)^{1/2}$ where $\widetilde{L_i}$ are scalar components of the chiral multiplets $L_i$ of $SU(2)_L$-doublet leptons. After integrating out the heavy Majorana neutrinos, $N_R$, the effective operator is induced at low energies
  \begin{eqnarray}
 W_{\rm eff}\supset\frac{1}{2{\cal M}_{i}}\,(\widetilde{L}_i\,H_u)^2\,,\qquad\text{with}~{\cal M}_{i}\equiv\frac{v^2_u}{(\hat{M}_{\nu\nu})_i}\,.
 \label{}
 \end{eqnarray}
 where $(\hat{M}_{\nu\nu})_i=(U^T_{\rm PMNS}M_{\nu\nu}U_{\rm PMNS})_{ii}\simeq\delta_i$ in Eq.\,(\ref{m_split0}).
Recalling that the $3\times3$ mixing matrix $U_L=U_{\rm PMNS}$ diagonalizing the mass matrix $M_{\nu\nu}=-m^T_D\,M^{-1}_R\,m_D$ participates in the charged weak interaction, the active neutrino mixing angles $(\theta_{12}, \theta_{13}, \theta_{23}, \delta_{CP})$ and the pseudo-Dirac mass splittings $\delta_k$ responsible for new wavelength oscillations characterized by the $\Delta m^2_k$ could be obtained from the mass matrix $M_{\nu\nu}$ formed by seesawing.  
Then, from Eqs.\,(\ref{Ynu1}) and (\ref{MR1})  we obtain the $\mu-\tau$ powered mass matrix as in Refs.\,\cite{Ahn:2012cg,Ahn:2014gva}
 \begin{eqnarray}
  M_{\nu\nu}
   &=& m_{0}\,e^{i\pi}
   {\left(\begin{array}{ccc}
   1+2F & (1-F)\,y_{2} & (1-F)\,y_{3} \\
   (1-F)\,y_{2} & (1+\frac{F+3\,G}{2})\,y^{2}_{2} & (1+\frac{F-3\,G}{2})\,y_{2}\,y_{3}  \\
   (1-F)\,y_{3} & (1+\frac{F-3\,G}{2})\,y_{2}\,y_{3} & (1+\frac{F+3\,G}{2})\,y^2_{3}
   \end{array}\right)}\nonumber\\
   &=&U^\ast_{\rm PMNS}\hat{M}_{\nu\nu}U^\dag_{\rm PMNS}\,,
\label{mass matrix}
 \end{eqnarray}
where
 \begin{eqnarray}
 m_{0}\equiv \left|\frac{\hat{y}^{\nu2}_{1}\upsilon^{2}_{u}}{3M}\right|\left(\frac{v_{T}}{\sqrt{2}\Lambda}\right)^2\left(\frac{v_{\Psi}}{\sqrt{2}\Lambda}\right)^{18},\quad F=\left(\tilde{\kappa}\,e^{i\phi}+1\right)^{-1},\quad G=\left(\tilde{\kappa}\,e^{i\phi}-1\right)^{-1}.
 \label{Numass1}
 \end{eqnarray}
In the limit  $y^\nu_1=y^\nu_2=y^\nu_3$ ($y_{2}, y_{3}\rightarrow1$), the  mass matrix\,(\ref{mass matrix}) gives the tri-bimaximal mixing (TBM) angles\,\cite{Harrison:2002er} and their corresponding mass eigenvalues $|\delta_k|$:
 \begin{eqnarray}
 &&~\qquad\sin^{2}\theta_{12}=\frac{1}{3}\,,\qquad\qquad\qquad\sin^{2}\theta_{23}=\frac{1}{2}\,,\qquad\qquad\qquad\sin\theta_{13}=0\,,\nonumber\\
 &&|\delta_1|=\frac{\Delta m^2_1}{2m_1}=3\,m_{0}\,|F|~,\qquad|\delta_2|=\frac{\Delta m^2_2}{2m_2}=3\,m_{0}\,,\qquad|\delta_3|=\frac{\Delta m^2_3}{2m_3}= 3\,m_{0}\,|G|~.
 \label{TBM1}
 \end{eqnarray}
These $|\delta_k|$ are disconnected from the TBM mixing angles. It is in general expected that deviations of $y_2, y_3$ from unity, leading to the non-zero reactor mixing angle\,\cite{An:2012eh, Ahn:2012nd}, {\it i.e.} $\theta_{13}\simeq8.5^{\circ}$ at $1\sigma$ best-fit\,\cite{Gonzalez-Garcia:2015qrr}, and in turn opening a possibility to search for CP violation in neutrino oscillation experiments.
These deviations generate relations between mixing angles and eigenvalues $|\delta_k|$.
Therefore Eq.\,(\ref{mass matrix}) directly indicates that there could be deviations from the exact TBM if the Dirac neutrino Yukawa couplings in $m_D$ of Eq.\,(\ref{Ynu1}) do not have the same magnitude, and the pseudo-Dirac mass splittings are all of the same order
 \begin{eqnarray}
 |\delta_1|\simeq |\delta_2|\simeq |\delta_3|\simeq {\cal O}(m_{0})\,.
 \label{m_split}
 \end{eqnarray}
As shown in Ref.\cite{Ahn:2016hbn} by numerical analysis, together with well-fitted $\theta_{12}$ and $\theta_{13}$ the values of atmospheric ($\theta_{23}$) and Dirac CP phase ($\delta_{CP}$) have a remarkable coincidence with the recent data by the NO$\nu$A\,\cite{Adamson:2017qqn} and/or T2K\,\cite{Abe:2017uxa} experiments.
From the overall scale of the mass matrix in Eq.\,(\ref{Numass1}) the pseudo-Dirac mass splitting, $\delta_2$, is expected to be
\begin{eqnarray}
  |\delta_2|\simeq2.94\times10^{-11}\left(\frac{4.24\times10^{9}{\rm GeV}}{M}\right)\left|\hat{y}^{\nu}_{1}\frac{v_{T}}{\sqrt{2}\Lambda}\right|^{2}\sin^2\beta~{\rm eV}\,,
 \label{scaleLambda}
\end{eqnarray}
in which the scale of the heavy neutrino, $M$, can be estimated from Eq.\,(\ref{MR2}) through the astrophysical constraints as $M=|\hat{y}_\Theta|\times2.75^{+1.50}_{-1.25}\times10^{9}\,{\rm GeV}$ which is connected to the PQ symmetry breaking scale via the axion decay constant in Ref.\,\cite{Ahn:2016hbn}. Eq.\,(\ref{scaleLambda}) shows that the value of $\delta_2$ depends on the magnitude $\hat{y}^{\nu}_{1}v_{T}/\Lambda$ since $M$ is constrained by the axion decay constraints: the smaller the ratio $v_{T}/\Lambda$, the smaller becomes $|\delta_k|$ responsible for the pseudo-Dirac mass splittings\,\footnote{Moreover, the overall scale of the heavy neutrino mass $M$ is closely related with a successful leptogenesis (see the details in Sec.\,\ref{AD}), constraints of the mass-squared differences in Eq.\,(\ref{msd}), and the CKM mixing parameters, therefore it is very important to fit the parameters $v_{T}/\Lambda$ and $M$.}. However, the value of $|\delta_k|$ is constrained from Eq.\,(\ref{D_lbound1}); for example, using $\tan\beta=2$ and $v_T/\Lambda\simeq\lambda^2/\sqrt{2}$ we obtain
\begin{eqnarray}
  |\delta_2|\simeq1.50\times10^{-14}\,|\hat{y}^\nu_1|^2\,{\rm eV}\,.
 \label{scaleLambda2}
\end{eqnarray}

 Since the potential is (almost) flat in these directions $\zeta_i$, they have large initial VEVs in the early universe, see Eq.\,(\ref{AD_min}). Such flat directions are lifted by some effective operators in a later epoch, receiving soft-masses in the SUSY breaking vacuum. 
Then the potential of the flat directions, $\zeta_i$, is directly written as
  \begin{eqnarray}
  V_0(\zeta_i)=m^2_{\zeta_i}|\zeta_i|^2+\frac{m_{3/2}}{8{\cal M}_{i}}(a_m\,\zeta^4_i+\text{h.c.})+\frac{|\zeta_i|^6}{4{\cal M}^2_{i}}\,.
 \label{AD_potential01}
 \end{eqnarray}
Here in the mass terms $m^2_{\zeta_i}$ we have included soft scalar masses generated by the $F$-term SUSY breaking, that is, the contribution from the effective $\mu$-term, $W\supset\mu_{\rm eff}\,H_uH_d$, which gives mass terms $\mu^2_{\rm eff}|\zeta_i|^2/2$. Since our model lies in the gravity-mediated SUSY breaking mechanism it is expected that $m_{\zeta_i}\sim m_{3/2}$ and $|a_m|\sim{\cal O}(1)$ in the $A$-term\,\footnote{In the context of Kallosh-Linde (KL) type models the dominant contributions to $A$-term arise from loop corrections\,\cite{Choi:2005ge} because at tree level $A$-terms are strongly suppressed by $m_{3/2}/m_T$, hence one needs relatively large ${\cal O}(100)$ TeV gravitino mass in order to get properly large $A$-terms\,\cite{Linde:2011ja}.}.
The potential for $\zeta_i$ in Eq.\,(\ref{AD_potential01}) is $D$-flat, $|\zeta_i|=0$, and also $F$-flat in the limit of $\delta_i ({\rm or}\,\Delta m^2_i)\rightarrow0$. So, the AD fields $\zeta_i$ can develop large VEVs during inflation. As discussed before, during inflation the energy density of the universe is dominated by the inflaton $\Psi_0$, that is, $V_0(t_I)=3H^2_I\,M^2_P$. 
The potential for $D$-flat direction is generated from the coupling between the AD fields $\zeta_i$ and the inflaton $\Psi_0$, which generically takes the form  
  \begin{eqnarray}
 K\supset K_{\rm AD}=|\Psi_0|^2+|\zeta_i|^2+\left(k_{\zeta i}\frac{|\Psi_0|^2}{M_{P}}\zeta_i+{\rm h.c.}\right)+\gamma_{\zeta i}\frac{|\Psi_0|^2|\zeta_i|^2}{M^2_P}+...\,,
 \label{Kahler_AD}
 \end{eqnarray}
where $k_{\zeta i}$ and $\gamma_{\zeta i}$ are complex and real constants, respectively, and the dots represent higher order terms which are irrelevant for our discussion.
Then, due to the finite energy density of the inflaton $\Psi_0$ during inflation the AD fields $\zeta_i$ receive additional SUSY breaking effects. And such SUGRA contribution reads
  \begin{eqnarray}
  V_{\rm sugra}(\zeta_i)=-\tilde{c}_HH^2_I|\zeta_i|^2+\frac{H_I}{8{\cal M}_{i}}(a_H\,\zeta^4_i+\text{h.c.})\,.
 \label{AD_potential02}
 \end{eqnarray}
Here by taking $\tilde{c}_H>0$ with $\tilde{c}_H$ being of order unity we assume that the AD fields $\zeta_i$ can obtain negative Hubble-induced mass terms.
From Eq.\,(\ref{AD_potential01}) and (\ref{AD_potential02}) the total effective potential for the AD fields $\zeta_i$ relevant to the leptogenesis reads
  \begin{eqnarray}
  V(\zeta_i)=V_0(\zeta_i)+V_{\rm sugra}(\zeta_i)\,.
 \label{AD_potential03}
 \end{eqnarray}
Then the minima of the potential are given by 
  \begin{eqnarray}
   \langle|\zeta_i|\rangle\simeq\left(\frac{4}{3}\tilde{c}_H\right)^{\frac{1}{4}}\left(\frac{m_i}{\Delta m^2_i}\,H_I\,v^2\sin^2\beta\right)^{\frac{1}{2}}\lesssim M_P\,,
 \label{AD_min}
 \end{eqnarray}
and $\arg(a_H)+4\arg(\zeta_i)\simeq\pi(2n+1)/2$ with $n=0,1$, in which we have used $m_{\zeta_i}, m_{3/2}|a_m|\ll H_I$.
The AD fields $\zeta_i$ at the origin are unstable due to the negative Hubble mass terms in Eq.\,(\ref{AD_potential02}), and so roll down toward their global SUSY minima of the potential in Eq.\,(\ref{AD_potential03}) during inflation. Thus, the AD fields $\zeta_i$ have large scales of $\sim\sqrt{v^2_u\,H_I/|\delta_i|}\lesssim M_P$ in Eq.\,(\ref{AD_min}) during inflation.
This is compatible with the fact that the Planck scale, $M_P$, sets the universe's minimum limit, beyond which the laws of physics break.
If we set the initial minima of the AD fields to the (almost) Planck scale, the ratios $m_i/\Delta m^2_i$ responsible for the neutrino mass splittings $\delta_i$ (relevant to the low energy neutrino oscillation as well as the high energy neutrino at the IceCube telescope) could be resricted as
  \begin{eqnarray}
   \frac{1}{\delta_i}=\frac{2\,m_i}{\Delta m^2_i}\lesssim\frac{M^2_P}{H_I\,v^2\sin^2\beta}\Big(\frac{3}{\tilde{c}_H}\Big)^{1/2}\,.
 \label{AD_ini}
 \end{eqnarray}
Using $H_I\simeq10^{10}$ GeV, $v=246$ GeV, $\sin\beta\simeq1$, and $1/\sqrt{10}\lesssim\tilde{c}_H\lesssim\sqrt{10}$, a lower bound can be roughly estimated as 
\begin{eqnarray}
 \delta_i\gtrsim(2-5)\times10^{-14}\,{\rm eV}
\end{eqnarray}
which is well compatible with the constraints from the neutrino data in Eq.\,(\ref{D_lbound1}) as well as a successful leptogenesis in Eq.\,(\ref{AD_6}).
 
After inflation ends, the inflaton $\Psi_0$ and waterfall field $\tilde{\Psi}$ (see Eqs.\,(\ref{mix_st}) and (\ref{reexp})) begin to oscillate around their VEVs, $\langle\tilde{\Psi}\rangle=\mu_{\Psi}$ and $\langle\Psi_0\rangle\simeq0$ (the VEV of $\Psi_0$ deviates from zero because of the supergravity effect: $\langle\Psi_0\rangle\sim m_{3/2}/|\tilde{g}_7|$ at the true minimum, see Eq.\,(\ref{ve0})) and their decays produce a dilute thermal plasma formed by collisions of relativistic decay products.
Since the energy density of the universe is still dominated by the inflaton $\Psi_0$ and waterfall field $\tilde{\Psi}$ during the inflaton and waterfall field oscillations epoch, the AD fields potential is still governed by the Hubble-induced mass terms in Eq.\,(\ref{AD_potential02}) together with $V_{0}(\zeta_i)$ in Eq.\,(\ref{AD_potential01}) at the first stage of oscillation. Thus, the AD fields $\zeta_i$ are trapped in the minima  determined mainly by the Hubble $A$-term as in Eq.\,(\ref{AD_min}) because the curvatures around the minima along both radial and angular directions are of the order of $H_I$ also in this period. 
However, after inflation the values of $\zeta_i$ in Eq.\,(\ref{AD_min}) gradually decrease to the order of $\zeta_i$ masses as the Hubble parameter $H(T)$ decreases, then the negative Hubble-induced mass terms are eventually exceeded by the Hubble parameter, {\it i.e.}, $\tilde{c}_HH(T)^2\lesssim m^2_{\zeta_i}$ in the potential Eq.\,(\ref{AD_potential03}). And the AD fields begin to oscillate around the potential minima $\langle\zeta_i\rangle\simeq0$ (actually, $m_{\zeta_i}$) with $H(T)=H_{\rm osc}$ when the Hubble parameter $H(T)$ of the universe becomes comparable to the SUSY breaking mass $m_{\zeta_i}$. (Hereafter ``osc" labels the epoch when the coherent oscillations commence.) Then the interactions of dimension-5 operators create lepton number. 

Now we see how the lepton number is created.
At the beginning of the oscillation, the AD fields have the initial values 
  \begin{eqnarray}
  |\zeta_i(t_{\rm osc})|\simeq\Big(\frac{4}{3}\tilde{c}_H\Big)^{1/4}\Big(\frac{m_{\zeta_i}\,m_i}{\Delta m^2_i}\,v^2\sin^2\beta\Big)^{1/2}\ll M_P\,.
    \label{AD_init}
 \end{eqnarray} 
in which $m_{\zeta_i}\simeq H_{\rm osc}$ is used.
The evolution of the AD fields $\zeta_i$ after $H\simeq H_{\rm osc}$ is described in a Friedmann-Robertson-Walker(FRW) universe by the equation of motion with the potential $V(\zeta_i)$ as
  \begin{eqnarray}
   \ddot{\zeta_i}+3H(T)\,\dot{\zeta_i}+\frac{\partial V(\zeta_i)}{\partial\zeta^\ast_i}\simeq0\,,
 \label{AD_eom}
 \end{eqnarray}
where $H(T)=(\pi^2g_{\ast}(T)/90M^{2}_{P})^{1/2}\,T^2\approx1.66\sqrt{8\pi\,g_\ast(T)}\,T^2/M_{P}$ is the Hubble rate for a radiation-dominated era with the total number of effective degrees of freedom $g_{\ast}(T)$ at a temperature $T$\,\cite{earlyUniverse}, $\partial V(\zeta_i)/\partial\zeta^\ast_i\simeq m^2_{\zeta_i}\,\zeta_i$, and dot indicates time derivative.
It is clear that the AD fields $\zeta_i$ oscillate around the origin ($\langle\zeta_i\rangle\simeq0$, the VEVs of $\zeta_i$ deviate from zero due to the SUGRA effect) and the amplitude of the oscillation damps as $|\zeta_i|\propto H\propto t^{-1}$.

Since the AD fields $\zeta_i$ carry lepton number, the baryon number asymmetry will be created during coherent oscillation of the AD fields. The number density of the AD fields is related to the lepton number density $n_{L_i}$ as $n_{L_i}=\frac{i}{2}(\frac{\partial\zeta^\ast_i}{\partial t}\zeta_i-\zeta^\ast_i\frac{\partial\zeta_i}{\partial t})$, then from Eq.\,(\ref{AD_eom}) the evolution of $n_{L_i}$ are given by
  \begin{eqnarray}
   \frac{\partial n_{L_i}}{\partial t}+3H\,n_{L_i}-\frac{m_{3/2}}{2{\cal M}_i}\,{\rm Im}(a_m\,\zeta^4_i)-\frac{H}{2{\cal M}_i}\,{\rm Im}(a_H\,\zeta^4_i)\simeq0\,.
\label{AD_1}
 \end{eqnarray}
Since the Hubble parameter $H(T)$ decreases as temperature decreases, the relative phase between $a_m$ and $a_H$ changes with time when the AD fields $\zeta_i$ trace the valleys determined mainly by the Hubble $A$-term\,\footnote{If there are no true minima, {\it i.e.} $m_{3/2}=0$, the AD fields get eternally trapped in the minima Eq.\,(\ref{AD_ini}) and there is no motion of $\zeta_i$ changing with time along the angular direction, leading to no lepton number production.}. And during their rolling towards the true minima, the contribution of ${\rm Im}(a_{H}\,\zeta_i^4)$ is suppressed compared with ${\rm Im}(a_{m}\,\zeta_i^4)$. Then the motion of $\zeta_i$ in the angular direction generating lepton number is expressed as
  \begin{eqnarray}
   \frac{\partial n_{L_i}}{\partial t}+3H\,n_{L_i}\simeq\frac{m_{3/2}}{2{\cal M}_i}\,{\rm Im}(a_m\,\zeta^4_i)\,,
\label{AD_2}
 \end{eqnarray}
 where $H=\dot{R}(t)/R(t)$,
and $R(t)$ stands for the scale factor of the expansion universe with cosmic time $t$.
The produced lepton number asymmetry at a time $t$ can be obtained by integrating the above equation $\partial(R^3\,n_{L_i})/\partial t\simeq\frac{m_{3/2}}{2{\cal M}_i}R^3\,{\rm Im}(a_m\,\zeta^4_i)$ where $R=R(t)$.
After the end of inflation, the inflaton field $\Psi_0$ and waterfall field $\tilde{\Psi}$ begin to oscillate around the potential minimum such that the universe is effectively matter dominated, which scales as $R^3\propto H^{-2}\propto t^2$. And  before the beginning of the $\zeta_i$ oscillation, due to $|\zeta_i|\propto H^{1/2}\propto t^{-1/2}$, the net lepton number generated keeps constant for the period $t<t_{\rm osc}$.
During matter dominated epoch the Hubble parameter is related to the expansion time by $H_{\rm osc}=(2/3)t^{-1}_{\rm osc}$. Then using Eq.\,(\ref{AD_init}) the generated lepton number at this stage $(t=t_{\rm osc})$ is given approximately by
  \begin{eqnarray}
   n_{L_i}(t_{\rm osc})\simeq\frac{\tilde{c}_H}{9}\,\frac{m_i\,v^2\sin^2\beta}{\Delta m^2_i}(m_{3/2}|a_m|)\,H_{\rm osc}\,\delta_{\rm eff}\,,
 \label{AD_3}
 \end{eqnarray}
where $\delta_{\rm eff}\simeq\sin(4\arg\zeta_i+\arg a_m)$ represents an effective $CP$ violating phase. It is expected that the production of net lepton asymmetry occurs before the reheating process completes, {\it i.e.}, $\Gamma_{\rm all}=\Gamma_{\Psi_0}+\Gamma_{\tilde{\Psi}}< H_{\rm osc}$, c.f., see Eq.\,(\ref{inf_reht}); the production of lepton number is strongly suppressed after the AD fields $\zeta_i$ start their oscillations, because ${\rm Im}(a_m\,\zeta^4_i)$ change their sign rapidly due to the oscillation of $\zeta_i$ as well as the amplitude of $\zeta_i$ oscillation is damped with expansion (see below Eq.\,(\ref{AD_eom})). Thus after inflation $R^3\,n_{L_i}\big|_{t=t_{\rm osc}}=R^3\,n_{L_i}\big|_{t=t_R}\sim n_{L_i}(t_R)/\rho_{\rm rad}(t_R)$ stays constant until the inflaton $\Psi_0$ and waterfall field $\tilde{\Psi}$ decays into light particles. Here $\rho_{\rm rad}(t_R)=3M^2_P\,\Gamma^2_{\rm all}$ is the energy density of the inflaton. 
Then the generated lepton number when the reheating process completes $(t=t_R,\,H\simeq\Gamma_{\rm all})$ is given by
  \begin{eqnarray}
   n_{L_i}(t_{R})= n_{L_i}(t_{\rm osc})\left(\frac{\Gamma_{\rm all}}{H_{\rm osc}}\right)^2\,.
 \label{AD_4}
 \end{eqnarray}
The inflaton decays reheats the universe producing entropy $s$ of radiation such that $\rho_{\rm rad}(t_R)=3T_{\rm reh}\,s(t_R)/4$. 
Then the lepton number asymmetry is approximately expressed as
  \begin{eqnarray}
   \frac{n_{L_i}(t_R)}{s}&=& \frac{\tilde{c}_H}{36}\frac{m_i\,v^2\sin^2\beta}{M^2_P\,\Delta m^2_i}\,T_{\rm reh}\left(\frac{m_{3/2}|a_m|}{H_{\rm osc}}\right)\delta_{\rm eff}
 \label{AD_5}
 \end{eqnarray}
when the reheating process of inflaton completes. Later, we will discuss the reheating temperature, see Sec.\,\ref{infret}, and its related gravitino problem, see Sec.\,\ref{infGraP}.
Recalling that the $H_{\rm osc}$ depends on ${\cal M}_i$ as $H_{\rm osc}\simeq m_{\zeta_i}$. Since ${\cal M}_i$ is directly related to the pseudo-Dirac mass splittings $\delta_i$ as ${\cal M}_i=\langle H_u\rangle^2/\delta_i$ in Eq.~(\ref{m_split0}) in addition to ${\cal O}(\delta_1)\simeq{\cal O}(\delta_2)\simeq{\cal O}(\delta_3)={\cal O}(m_0)$ in Eq.~(\ref{m_split}), there are three flat directions corresponding to the almost degenerate neutrino pairs, 
{\it i.e.}, the three generation AD fields $\zeta_i/\sqrt{2}=\widetilde{L}_i=H_u$ with $i=1,2,3$. 
The lepton asymmetries in Eq.\,(\ref{AD_5}) are converted into the baryon asymmetry through non-perturbative sphaleron processes. We are in the energy scale where $A_4\times U(1)_X\times{\rm SUSY}$ is broken but the SM gauge group remains unbroken. So the baryon number produced is thermalized in a hot plasma into real baryons at a relatively low temperature. 
Therefore, the present baryon asymmetry can be expressed by
  \begin{eqnarray}
   \frac{n_B}{s}&\simeq&0.35\sum_{i=1,2,3}\frac{n_{L_i}}{s}\nonumber\\
   &\simeq&8.67\times10^{-11}\times\frac{\sum_{i=1}^3\frac{ m_i}{\Delta m^2_i}}{1.75\times10^{10}\,{\rm eV}^{-1}}\left(\frac{T_{\rm reh}}{10^3\,{\rm TeV}}\right)\left(\frac{\delta_{\rm eff}}{0.1}\right)\left(\frac{\tilde{c}_{H}}{0.5}\right)\left(\frac{m_{3/2}|a_m|}{H_{\rm osc}}\right)\,,
 \label{AD_6}
 \end{eqnarray}
where $n_B$ is the baryon number density and $s$ is the entropy density, and we have used $\sin\beta\simeq1$. Considering $1/\sqrt{10}\lesssim|a_m|, \tilde{c}_H\lesssim\sqrt{10}$ (being order of unity) and $H_{\rm osc}\simeq m_{3/2}\simeq m_{\zeta_i}$\,\footnote{Recalling that our scenario lies in the gravity-mediated SUSY breaking mechanism, see below Eq.\,(\ref{AD_potential01}).} and, for convenience, defining $x_{\rm reh}\equiv(m_{3/2}|a_m|/H_{\rm osc})\delta_{\rm eff}\,\tilde{c}_H$, the resultant baryon asymmetry only depends on the neutrino parameters $m_i$ and $\Delta m^2_i$, $T_{\rm reh}$, and $x_{\rm reh}$.
\begin{figure}[h]
\includegraphics[width=8.0cm]{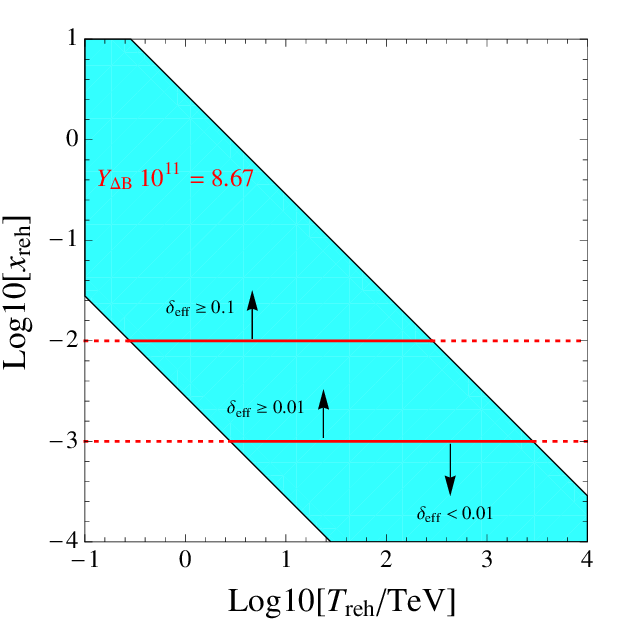}
\caption{\label{Fig2} Regionplot for the successful leptogenesis $Y_{\Delta_B}=8.67\times10^{-11}$ (cyan region) as a function of $T_{\rm reh}/{\rm TeV}$ and $x_{\rm reh}\equiv(m_{3/2}|a_m|/H_{\rm osc})\delta_{\rm eff}\,\tilde{c}_H$ where the regions $10^{10}\,\text{eV}^{-1}\lesssim\sum_{i}\frac{m_{\nu_i}}{\Delta m^2_i}\lesssim5\times10^{13}\,\text{eV}^{-1}$ in Eq.\,(\ref{nu_sum_d}) is used. Especially, for the case of $m_{3/2}\simeq H_{\rm osc}$, $1/\sqrt{10}\lesssim\tilde{c}_H, |a_m|\lesssim\sqrt{10}$, and $\delta_{\rm eff}\leq1$, the horizontal line represents a lower bound of $x_{\rm reh}$.}
\end{figure}
 Once the values of $T_{\rm reh}$ and $x_{\rm reh}$ are fixed, quantitatively, the value of BAU is inferred from the two observations, $m_i$ ($\simeq m_{\nu_i}$) and $\Delta m^2_i$, independently: from Eqs.\,(\ref{D_lbound1}), (\ref{nu_sum_c}), and (\ref{AD_ini}) the following quantity could be extracted as 
 \begin{eqnarray}
 10^{10}\,\text{eV}^{-1}\lesssim\sum_{i}\frac{m_{\nu_i}}{\Delta m^2_i}=\frac{1}{2}\left(\frac{1}{\delta_1}+\frac{1}{\delta_2}+\frac{1}{\delta_3}\right)\lesssim5\times10^{13}\,\text{eV}^{-1}\,,
  \label{nu_sum_d}
 \end{eqnarray}
in which the upper bound is derived from an initial condition of the AD fields in Eq.\,(\ref{AD_ini}); the lower bound comes from the neutrino data in Eqs.\,(\ref{D_lbound1}) and (\ref{nu_sum_c}).
In terms of $Y_{\Delta B}\equiv(n_B-n_{\bar{B}})/s|_{\rm today}$ (which is conserved throughout the thermal evolution of the universe) the BBN results\,\cite{Iocco:2008va} and the CMB measurement\,\cite{Ade:2015xua} read at $95\%$ CL
  \begin{eqnarray}
   Y^{\rm BBN}_{\Delta B}=(8.10\pm0.85)\times10^{-11}\,,\qquad  Y^{\rm CMB}_{\Delta B}=(8.67\pm0.05)\times10^{-11}\,.
 \label{observe}
 \end{eqnarray}
As shown in Fig.\,\ref{Fig2}, taking into account $\delta_{\rm eff}\geq0.01$ (see, below Eq.\,(\ref{AD_3})), $1/\sqrt{10}\lesssim\tilde{c}_H, |a_m|\lesssim\sqrt{10}$ (see, below Eqs.\,(\ref{AD_potential02}) and (\ref{AD_potential01})), and $10^{10}\,\text{eV}^{-1}\lesssim\sum_{i}\frac{m_{\nu_i}}{\Delta m^2_i}\lesssim5\times10^{13}\,\text{eV}^{-1}$ in Eq.\,(\ref{nu_sum_d}), for the baryon asymmetry in Eq.\,(\ref{AD_6}) to satisfy the BBN results and CMB measurement a range of plausible reheating temperature could be obtained as
  \begin{eqnarray}
   {\cal O}(100)\,{\rm GeV}\lesssim T_{\rm reh}\lesssim3\times10^3\,\text{TeV}\,,
 \label{Tr_pre}
 \end{eqnarray}
 where the lower bound is due to electroweak scale.
Later, we will show that the bound of Eq.\,(\ref{Tr_pre}) could be consistent with the bound from Eq.\,(\ref{let_reh}).

\subsection{Gravitino production}
\label{infGraP}
It is well known that thermal leptogensis in supersymmetric framework, which is one of attractive mechanism for origin of matter, requires a large reheating temperature in the early universe, $T_{\rm reh}\sim M_1>10^{9}$ GeV, where $M_1$ is a lightest heavy neutrino mass. The gravitino, which appears in all models with local supersymmetry, is the superpartner of the graviton. Gravitino is produced thermally\,\cite{Bolz:2000fu} or non-thermally\,\cite{Kallosh:1999jj, Nilles:2001ry, Endo:2006zj, Kawasaki:2006gs, Endo:2007sz} in the cosmological history. The excessive production of gravitinos in the early universe may destroy the nucleosynthesis of the light elements for unstable gravitinos or overclose of universe for stable gravitinos\,\cite{myk}.
Since the gravitino is present in the supersymmetric model, we are going to address (unstable) gravitino overabundance problem.

As mentioned in Sec.\,\ref{A4U1}, there are two secluded SUSY breaking sectors, {\it i.e.}, SUSY$=$SUSY$_{\rm inf}\times$SUSY$_{\rm vis}$. 
Gravitational interactions explicitly break the SUSY down to {\it true} SUSY$_{\rm inf}\times$SUSY$_{\rm vis}$, where SUSY$_{\rm inf}$ corresponds to the genuine SUGRA symmetry, while the orthogonal SUSY$_{\rm vis}$ is approximate global symmetry. In each sector, spontaneous breakdown of $F$-term occurs at a scale $F_i$ ($i=$ inf, vis) independently, producing a corresponding goldstino. Hence, in the presence of SUGRA, the SUSY$_{\rm inf}$ is gauged and thus its corresponding goldstino is eaten by the gravitino via super-Higgs mechanism, leaving behind the approximate global symmetry SUSY$_{\rm vis}$ which is explicitly broken by SUGRA and thus its corresponding the uneaten goldstino as a propagating degree of freedom.

During inflation and the beginning of reheating (preheating) when SUSY is spontaneously broken there are possible productions of fermonic quanta which are strongly coupled to the inflaton field.
During this stage the SUSY$_{\rm inf}$ is mainly broken by the inflaton implying that the goldstino produced is mainly inflatino (instead of the gravitino in the low energy); the gravitino produced non-thermally\,\footnote{The inflatinos produced during inflation and preheating may be partially converted to the gravitinos in the low energy, since $G_{\Psi_0}$ is generically non-zero in the true minimum\,\cite{Nakamura:2006uc}. At this stage, since the inflationary sector and the sector responsible for the low energy effective SUSY breaking are distinct, the gravitinos generated non-thermally are produced with a sufficiently low abundance.} is effectively massless as long as the Hubble parameter is larger than the gravitino mass, $H>m_{3/2}$\,\cite{Nilles:2001ry}. However, this correspondence does not necessarily hold at late times, since the SUSY$_{\rm vis}$ is broken by other fields in the true vacuum.

After the inflation ends, the inflaton $\Psi_0$ and waterfall field $\tilde{\Psi}$ release their energy into a thermal plasma by the decays, and the universe is reheated. Since all the particles including photons and baryons in the present universe are ultimately originated from the decays, it is crucial to reveal how the reheating proceeds. 
In SUGRA framework, with the linear Kahler potential in Eq.\,(\ref{NK}) the inflaton field $\Psi_0$ has a non-vanishing auxiliary field $G_{\Psi_0}$. Such non-vanishing auxiliary field allows the inflaton decay into a pair of the gravitinos, whose decay process is crucial in the reheating process\,\cite{Kawasaki:2006gs}.
The constraint on the inflaton potential $G_{\Psi_0}$ depending on the gravitino mass must be satisfied to avoid an overproduction of the gravitino keeping the success of the standard cosmology.
In the unitary gauge in the Einstein frame, the goldstino (the longitudinal component of the gravitino) can be gauged away through the super-Higgs mechanism leading to vanishing of the gravitino-goldstino mixing. Then the relevant interactions for the inflaton decay into a pair of gravitinos reads\,\cite{wbg}
\begin{eqnarray}
 -e^{-1}{\cal L}&=&\frac{1}{8}\epsilon^{\mu\nu\rho\sigma}\left(G_{\Psi_0}\partial_\rho\Psi_0-G_{\bar{\Psi}_0}\partial_\rho\Psi^\ast_0\right)\bar{\psi}_\mu\gamma_\nu\psi_\sigma\nonumber\\
 &+&\frac{e^{G/2}}{8}M_P\left(G_{\Psi_0}\Psi_0+G_{\bar{\Psi}_0}\Psi^\ast_0\right)\bar{\psi}_\mu[\gamma^\mu,\gamma^\nu]\psi_\nu
 \label{gra_int}
\end{eqnarray}
where $\psi_\mu$ is the gravitino field.
The real and imaginary components of the inflaton field have the same decay rate at leading order\,\cite{Endo:2006zj}
\begin{eqnarray}
 \Gamma_{3/2}\equiv\Gamma(\Psi_0\rightarrow\psi_{3/2}+\psi_{3/2})\simeq\frac{1}{288\pi}\frac{M^2_P}{K_{\Psi_0\bar{\Psi}_0}}\big|\langle G_{\Psi_0}\rangle\big|^2\Big(\frac{m_{\Psi_0}}{M_P}\Big)^2\Big(\frac{m_{\Psi_0}}{m_{3/2}}\Big)^2m_{\Psi_0}
 \label{gam32}
\end{eqnarray}
in the limit of $m_{\Psi_0}\gg m_{3/2}$ after canonical normalization $\hat{\Psi}_0=\sqrt{K_{\Psi_0\bar{\Psi}_0}}\Psi_0$. The decay rate is enhanced by the gravitino mass in the denominator, which comes from the goldstino (mainly as the inflatino) in the massless limit.
The decay into the gravitinos only proceeds at the stage $H<m_{3/2}$, when the SUSY breaking contribution of the inflaton is subdominant\,\cite{Kawasaki:2006gs}. Thus, the gravitinos produced at the reheating epoch by the inflaton decay through the interaction (\ref{gra_int}) should coincide with those in the low energy.

Now, we estimate how much the gravitinos are produced at the reheating epoch. After the inflation ends both the inflaton $\Psi_0$ and waterfall field $\tilde{\Psi}$ oscillate around the potential minimum and dominate the universe until the reheating. 
Using $|G_{\Psi_0}|\lesssim|\Psi_0|/M^2_P$ one obtains $W_{\Psi_0}/W\simeq\Psi_0/M^2_P$.
Inserting $G_{\Psi_0\Psi_0}=-W^2_{\Psi_0}/W^2$, $G_{\Psi\Psi_0}\simeq-\Psi\,{W_{\Psi_0}}/(WM^2_P)\pm\tilde{g}_7\tilde{\Psi}/(m_{3/2}M^2_P)$, and $G_{z\Psi_0}\simeq\sqrt{3}W_{\Psi_0}/(WM_P)$ into Eqs.\,(\ref{mcon1}) and (\ref{mcon2}) we obtain
\begin{eqnarray}
 \langle{G}_{\Psi_0}\rangle\sim\frac{3\langle\Psi_0\rangle}{M^2_P}\simeq3\frac{m_{3/2}}{|\tilde{g}_7|M^2_P}\,,\qquad
 \langle{G}_{\Psi}\rangle\sim\frac{3}{2}\frac{m^2_{3/2}}{|\tilde{g}_7|^2}\frac{\langle\Psi\rangle}{M^4_P}\,,
\end{eqnarray}
which indicates $\langle{G}_{\Psi_0}\rangle$ is much larger than $\langle{G}_{\Psi}\rangle$. Then, from Eqs.\,(\ref{gam32}) and (\ref{infl_ma1}) the inflaton decay width is roughly given by
\begin{eqnarray}
 \Gamma_{3/2}\simeq\frac{1}{32\pi}\Big(\frac{m_{\Psi_0}}{M_P}\Big)^4\Big(\frac{\mu_{\Psi}(t_I)}{M_P}\Big)^2m_{\Psi_0}\,.
\label{ga32}
\end{eqnarray}

At the reheating epoch, gravitinos are produced by the non-thermal inflaton decay process ($Y^{\Psi_0}_{3/2}$: the yield of the gravitinos by the inflaton decay) as well as by the thermal scattering ($Y^{\rm th}_{3/2}$: the yield of the gravitinos produced by thermal scatterings);  the ratio of gravitino-to-entropy density is given by $Y_{3/2}=Y^{\Psi_0}_{3/2}+Y^{\rm th}_{3/2}$, which remains constant as the universe expands as long as there is no additional entropy production. Gravitinos\,\footnote{The production of gravitinos after inflation has been studied in some detail\,\cite{Khlopov:1984pf}.} thermally produced in the early universe, predominantly via $2\rightarrow2$ inelastic scatterings of gluons and gluinos by QCD process, have a potential problem for the thermal history of the universe. 
 However, since their relic density, $\Omega^{\rm th}_{3/2}h^2$, and contribution to the energy density, $Y^{\rm th}_{3/2}$, grow with the reheating temperature after inflation, the yield of the gravitinos thermally produced is estimated as $Y^{\rm th}_{3/2}\sim10^{-16}\,(T_{\rm reh}/10^3\,{\rm TeV})$\,\cite{Bolz:2000fu, Kawasaki:2004qu} which is harmless with the gravitino mass $m_{3/2}\sim100$ TeV in Eq.\,(\ref{gravi32}) with the reheating temperature satisfying the successful leptogenesis in Eq.\,(\ref{Tr_pre}).
On the other hand, the gravitino yield produced by the inflaton decay process $\Psi_0\rightarrow\Psi_{3/2}+\Psi_{3/2}$ via the interaction Eq.\,(\ref{gra_int}) is 
\begin{eqnarray}
  Y^{\Psi_0}_{3/2}\equiv\frac{n^{\Psi_0}_{3/2}}{s}\simeq2\frac{\Gamma_{3/2}}{\Gamma_{\Psi_0}}\frac{3}{4}\frac{T_{\rm reh}}{m_{\Psi_0}}\,,
  \label{y32t}
\end{eqnarray}
where $n^{\Psi_0}_{3/2}$ is the number density of gravitinos by the inflaton decay, and $s=(2\pi^2/45)g_{\ast s}(T)\,T^3$ is the entropy density with $g_{\ast s}(T)$ being the effective number of the massless degrees of freedom at the temperature $T$. 

The gravitino yield is severely constrained by BBN, $Y_{3/2}<Y^{\rm BBN}_{3/2}$,
in order to keep the success of the standard scenario of BBN\,\cite{Khlopov:1984pf}. Otherwise, the decay products of the gravitino would change the abundances of primordial light elements too much and consequently conflict with the observational data.
Refs.\,\cite{Kohri:2005wn, Ellis:1984er} shows that, when the hadronic branching ratio of the gravitino decay is of order unity, $Y^{\rm BBN}_{3/2}\sim10^{-16}$ for $m_{3/2}\sim1$ TeV and $Y^{\rm BBN}_{3/2}\sim10^{-15-13}$ for $m_{3/2}\sim10$ TeV; for $m_{3/2}\gtrsim100$ TeV the constraint disappears. 
On the other hand, in the context of supersymmetric moduli stabilization where moduli are strongly stabilized, at tree level the gaugino masses and $A$-terms are strongly suppressed by $m_{3/2}/m_{T}$ and as such effectively vanish\,\cite{Linde:2011ja}, while the dominant contributions to the gaugino masses and $A$ terms arise from loop corrections\,\cite{Choi:2005ge}: $m_{1/2}=b_ag^2_a/(16\pi^2)(F^C/C_0)$ and $A_{ijk}=-(\gamma_{ijk}/16\pi^2)(F^C/C_0)$ where $b_a=11,1,-3$ for $a=1,2,3$ are the one-loop beta function coefficients, $\gamma_{ijk}$ are the anomalous dimensions of the matter fields, and $F^C/C_0\sim m_{3/2}$. Thus, in order to have suitably large gaugino masses, relatively large ${\cal O}(100)$ TeV gravitino masses must be considered\,\cite{Linde:2011ja}.

\subsection{Reheating temperature}
\label{infret}
In order to estimate $Y^{\Psi_0}_{3/2}$ we have to calculate the decay width of the inflaton and waterfall fields, $\Gamma_{\rm all}$, at reheating epoch.

Since inflation leaves the early universe cold and empty, the inflaton $\Psi_0$ and waterfall field $\tilde{\Psi}$ where all energy resides in must transfer their energy to a radiation dominated plasma in local thermodynamic equilibrium at a temperature sufficient to allow standard nucleosynthesis $T_{\rm reh}>T({\rm BBN})$. So the universe must be reheated after inflation. 
The energy of the inflaton $\Psi_0$ and waterfall field $\tilde{\Psi}$ are transferred to the SM sector through their gravitational and/or non-gravitational decays once their fields acquire finite VEVs, which in turn produce SM matter. Their decay products thermalize.

We are in the case where the inflaton $\Psi_0$ and waterfall field $\tilde{\Psi}$ dominate the energy of the universe when they decay.
The reheating temperature $T_{\rm reh}$ resulting from the perturbative decays of the inflaton $\Psi_0$ and waterfall field $\tilde{\Psi}$\,\footnote{The energy transfer from the inflaton and waterfall field to the SM fields in general proceeds both through non-perturbative effects and perturbative decays\,\cite{Kofman:1997yn}} may be estimated by using the relation 
 \begin{eqnarray}
 \Gamma_{\rm all}=3H(T_{\rm reh})
 \label{tot_dec}
 \end{eqnarray}
at the end of the reheating process, where the Hubble parameter $H(T)$ is given in the radiation dominated era of the universe.
Inflaton $\Psi_0$ and waterfall field $\tilde{\Psi}$ decays reheat the universe, when $\Gamma_{\rm all}\gtrsim3H(T_{\rm reh})$:
   \begin{eqnarray}
 T_{\rm reh}=\left(\frac{10}{\pi^2 g_\ast}\right)^{1/4}\sqrt{\Gamma_{\rm all}M_{P}}\,,\qquad\text{with}~\Gamma_{\rm all}=\Gamma^{{\rm sugra}}_{\Psi_0}+\Gamma^{{\rm sugra}}_{\tilde{\Psi}}+\Gamma^{{\rm vis}}_{\Psi_0}+\Gamma^{{\rm vis}}_{\tilde{\Psi}}
  \label{inf_reht}
 \end{eqnarray}
 where $g_\ast(T)$ is the number of the relativistic degrees of freedom in the plasma\,\footnote{We estimate the total number of effectively massless degree of freedom of the radiation, $g_\ast(T)$, at temperature of the order of the decay rate of the inflaton and waterfall field $\Gamma_{\rm all}$, i.e., there are 17 bosons and 48 Weyl fermions for $T_{\rm EW}<T<m_{3/2}$: $g_\ast(T)=\sum_{j={\rm bosons}}g_j(T_j/T)^4+(7/8)\sum_{j={\rm fermions}}g_j(T_j/T)^4=34+(7/8)96=118$ where $T_j$ denotes the effective temperature of any species $j$.}, and $\Gamma^{{\rm sugra}}_{\Psi_0}+\Gamma^{{\rm sugra}}_{\tilde{\Psi}}$ and $\Gamma^{{\rm vis}}_{\Psi_0}+\Gamma^{{\rm vis}}_{\tilde{\Psi}}$ stand for gravitational and non-gravitational decay widths, respectively. 

As in Ref.\,\cite{Ahn:2016hbn}, in the supersymmetric visible sector the inflaton $\Psi_0$ and waterfall field $\tilde{\Psi}$ couple to the SM particles via the following interactions dominantly
  \begin{eqnarray}
  W\supset g_{\Psi_0}\Psi_0 H_uH_d+\hat{y}_c\Big(\frac{\tilde{\Psi}}{\Lambda}\Big)^2Q_2c^cH_u
  \label{nong}
 \end{eqnarray}
 where $g_{\Psi_0}$ is a real and positive coupling constant, while the hat Yukawa coupling $\hat{y}_c$ is of order unity complex number. Here $Q_2$ is the second generation left handed quark doublet, which transforms as ${\bf 1}''$ under $A_4$ symmetry; the right handed charm quark $c^c\sim{\bf 1}'$ under $A_4$.  
The first term is also associated with the $\mu$-term since the VEV of $\Psi_0$ is given by $\langle\Psi_0\rangle\sim m_{3/2}/|\tilde{g}_7|$. 
And so the inflaton with a non-zero VEV can decay into the visible sector through the non-gravitational coupling of the inflaton to matter with the decay rate
 \begin{eqnarray}
 \Gamma^{\rm vis}_{\Psi_0}&=&\Gamma(\Psi_0\rightarrow2\text{ Higgsinos})+\Gamma(\Psi_0\rightarrow2\text{ Higgses})\nonumber\\
 &\simeq&2\times\frac{|g_{\Psi_0}|^2}{16\pi}m_{\Psi_0}\,,
  \label{dec_psi0}
\end{eqnarray}
where the masses of the final-states compared to that of the inflaton are neglected. 
For the second term in Eq.\,(\ref{nong}), expanding the waterfall field $\tilde{\Psi}$ and the Higgs field $H_u$, without loss of generality, as
\begin{eqnarray}
 \tilde{\Psi}=\frac{1}{\sqrt{2}}\Big(v_{\tilde{\Psi}}+\frac{h_{\tilde{\Psi}}}{\sqrt{2}}-i\frac{\phi_\Psi}{\sqrt{2}}\Big)\,,\qquad H_u={\left(\begin{array}{c}
   v_u+\frac{h_u}{\sqrt{2}} \\
   0
   \end{array}\right)}\,,
\end{eqnarray}
the second term in Eq.\,(\ref{nong}) is expressed in terms of Lagrangian form as
 \begin{eqnarray}
  -{\cal L}=\hat{y}_c\Big(\frac{v_{\tilde{\Psi}}}{\sqrt{2}\Lambda}\Big)^2v_u\Big\{1+\frac{h_u}{\sqrt{2}v_u}+\frac{\sqrt{2}}{v_{\tilde{\Psi}}}(h_{\tilde{\Psi}}-i\phi_\Psi)\Big\}\bar{c}_Lc_R+{\rm h.c.}.
\end{eqnarray}
Here the waterfall field $\tilde{\Psi}$ with a non-zero VEV can decay into the visible sector through the non-gravitational coupling of the waterfall field $\tilde{\Psi}$ to matter with the decay rate
 \begin{eqnarray}
 \Gamma^{\rm vis}_{\tilde{\Psi}}\simeq\Gamma(\tilde{\Psi}\rightarrow c\bar{c})&\simeq& \frac{|\hat{y}_c|^2}{8\pi}\Big(\frac{v_{\tilde{\Psi}}}{\sqrt{2}\Lambda}\Big)^4\Big(\frac{v_u}{v_{\tilde{\Psi}}}\Big)^2m_{\tilde{\Psi}}\nonumber\\
 &=&\frac{|g_{\tilde{\Psi}}|^2}{8\pi}m_{\tilde{\Psi}}\,,
 \label{dec_psi}
\end{eqnarray}
where $g_{\tilde{\Psi}}\equiv\hat{y}_c(v_{\tilde{\Psi}}/\sqrt{2}\Lambda)^2(v_u/v_{\tilde{\Psi}})$, and the mass of the final-state compared to that of the waterfall field $\tilde{\Psi}$ is neglected. Using $|\hat{y}_c|\simeq1$, $v_{\tilde{\Psi}}/\sqrt{2}\Lambda=\lambda/\sqrt{2}$ and $v_{u}/v_{\tilde{\Psi}}\simeq10^{-8}$ where $\lambda\approx0.225$, $\sin\beta\simeq1$ and $v_{\tilde{\Psi}}\approx1.7\times10^{10}$ GeV\,\cite{Ahn:2016hbn}, we obtain
 \begin{eqnarray}
  |g_{\tilde{\Psi}}|\simeq2.5\times10^{-10}\,.
\end{eqnarray}

Next, we consider the gravitational effects on the reheating temperature. 
The inflaton $\Psi_0$ and waterfall field $\tilde{\Psi}$ with non-zero VEVs can also decay into the visible sector through the SUGRA effects\,\cite{Endo:2007sz}. Then the reheating can be induced by the inflaton and waterfall fields decay through non-renormalizable interactions. The relevant interactions for the matter-fermion production are provided in the Einstein frame as\,\cite{wbg}
\begin{eqnarray}
 e^{-1}{\cal L}&=&\frac{i}{2}K_{ij^\ast}\bar{\chi}^j\gamma^\mu\partial_\mu\chi^i+\frac{i}{8M^2_P}K_{ij^\ast}\left(K_\sigma\partial_\mu\phi^\sigma-K_{\sigma^\ast}\partial_\mu\phi^{\ast\sigma}\right)\bar{\chi}^j\gamma^\mu\chi^i\nonumber\\
 &-&\frac{i}{2M_P}K_{ij^\ast}\Gamma^i_{\sigma\rho}(\partial^\mu\phi^\sigma)\bar{\chi}^j\gamma^\mu\chi^\sigma+\frac{1}{2}e^{K/2M^2_P}({\cal D}_iD_jW)\chi^i\chi^j+h.c.
\label{suglg}
\end{eqnarray}
where ${\cal D}_iD_jW=W_{ij}+\frac{K_{ij}}{M^2_P}W+\frac{K_i}{M^2_P}D_jW+\frac{K_j}{M^2_P}D_iW-\frac{K_iK_j}{M^4_P}W-\frac{\Gamma^k_{ij}}{M_P}D_kW$. Here $\phi^i$ and $\chi^i$ stand for the matter fields, and $\phi^i$ collectively denotes on arbitrary fields including the inflaton $\Psi_0$ and waterfall field $\tilde{\Psi}$.
And the matter-scalar production is represented by the kinetic term and the scalar potential
\begin{eqnarray}
 -e^{-1}{\cal L}&=&iK_{ij^\ast}\partial_\mu\phi^i\partial^\mu\phi^{\ast j}+e^{K/M^2_P}\left\{K^{ij^\ast}(D_iW)(D_{\bar{j}}\bar{W})-\frac{3}{M^2_P}|W|^2\right\}\,.
\end{eqnarray}
In the model superpotential the supersymmetric visible sector contains the following renormalizable interactions  
\begin{eqnarray}
 W\supset y_t\,Q_3t^c\,H_u+\frac{1}{2}M_RN^cN^c\,,
\end{eqnarray}
where the first term is the top quark operator as in\,\cite{Ahn:2014gva} and the second term comes from Eq.\,(\ref{Axion_nu_La}) after the $U(1)_X$ is spontaneously broken. 
First, we consider the partial decay width of the inflaton. The partial decay width of the inflaton through the neutrino Yukawa coupling is\,\cite{Endo:2007sz}
\begin{eqnarray}
 \Gamma^{N({\rm sugra})}_{\Psi_0}&=&\Gamma(\Psi_0\rightarrow N^cN^c)+\Gamma(\Psi_0\rightarrow \tilde{N}^c\tilde{N}^c)\nonumber\\
 &\simeq&2\times\frac{c^N_{\Psi_0}}{32\pi}m_{\Psi_0}\left(1-\frac{4M^2}{m^2_{\Psi_0}}\right)^{1/2}\,,
 \label{gr1}
\end{eqnarray}
where $c^N_{\Psi_0}\simeq{e}^{K/M^2_P}\left|\frac{K_{\Psi_0}}{M^2_P}W_{N^cN^c}-2\Gamma^k_{\Psi_0N^c}\frac{W_{N^ck}}{M_P}\right|^2$; (sum over $k$) and the heavy neutrino mass $M$ given in Eq.\,(\ref{MR2}). For the minimal Kahler potential, for simplicity, using Eq.\,(\ref{ve0}) the parameter $c^N_{\Psi_0}$ can be approximately given by
\begin{eqnarray}
   c^N_{\Psi_0}\simeq\left(\frac{\langle\Psi_0\rangle}{M_P}\right)^2\left(\frac{M}{M_P}\right)^2=\left(\frac{m_{3/2}}{m_{\Psi_0}}\right)^2\left(\frac{\mu_\Psi(t_I)}{M_P}\right)^2\left(\frac{M}{M_P}\right)^2\,,
\end{eqnarray}
where in the last equality the inflaton mass $m_{\Psi_0}$  in Eq.\,(\ref{inf_ma}) or Eq.\,(\ref{infl_ma1}) is used. And the partial decay width of the inflaton through the top quark Yukawa coupling is\,\cite{Endo:2007sz}
\begin{eqnarray}
 \Gamma^{t({\rm sugra})}_{\Psi_0}&=&\Gamma(\Psi_0\rightarrow 3\,\text{scalars})+\Gamma(\Psi_0\rightarrow 1\,\text{scalar}+2\,\text{fermions})\nonumber\\
 &\simeq&\frac{c^t_{\Psi_0}\,6}{256\pi^3}\left(\frac{m_{\Psi_0}}{M_P}\right)^2m_{\Psi_0}\,,
 \label{gr2}
\end{eqnarray}
where the masses of the final state particles are neglected, the additional numerical factor comes from $SU(3)\times SU(2)$, and $c^t_{\Psi_0}\simeq{e}^{K/M^2_P}\left|\frac{K_{\Psi_0}}{M_P}W_{t^cQ_3H_u}-3\Gamma^\ell_{\Psi_0H_u}W_{t^cQ_3\ell}\right|^2$; (sum over $\ell$). Similarly, the parameter $c^t_{\Psi_0}$ is approximately given by
\begin{eqnarray}
 c^t_{\Psi_0}\simeq\left(\frac{\langle\Psi_0\rangle}{M_P}\right)^2|y_t|^2=\left(\frac{m_{3/2}}{m_{\Psi_0}}\right)^2\left(\frac{\mu_\Psi(t_I)}{M_P}\right)^2|y_t|^2\,.
 \end{eqnarray}
In addition, the decay rate into the visible sector through the top and neutrino Yukawa couplings is much larger than that into the gluons and gluoinos via the anomalies of SUGRA\,\cite{Endo:2007sz}.
Then, from Eqs.\,(\ref{gr1}) and (\ref{gr2}) the inflaton decay rate through the gravitational coupling of the inflaton to matter is approximately given by
\begin{eqnarray}
\Gamma^{{\rm sugra}}_{\Psi_0}&\simeq&\Gamma^{t({\rm sugra})}_{\Psi_0}+\Gamma^{N({\rm sugra})}_{\Psi_0}\nonumber\\
&\simeq& \frac{m_{\Psi_0}}{16\pi}\Big(\frac{m_{3/2}}{m_{\Psi_0}}\Big)^2\Big(\frac{\mu_\Psi(t_I)}{M_P}\Big)^2\Big\{\frac{2|y_t|^2}{8\pi^2}\Big(\frac{m_{\Psi_0}}{M_P}\Big)^2+\Big(\frac{M}{M_P}\Big)^2\Big(1-\frac{4M^2}{m^2_{\Psi_0}}\Big)^{\frac{1}{2}}\Big\}\,.
 \label{gr11}
\end{eqnarray}
Second, similar to the above case of the inflaton field, 
the waterfall field decay rate through the gravitational coupling of the waterfall field to matter is approximately given by
\begin{eqnarray}
\Gamma^{{\rm sugra}}_{\tilde{\Psi}}&\simeq&\Gamma^{t({\rm sugra})}_{\tilde{\Psi}}+\Gamma^{N({\rm sugra})}_{\tilde{\Psi}}\nonumber\\
&\simeq& \frac{m_{\tilde{\Psi}}}{16\pi}\Big(\frac{\mu_\Psi(t_I)}{M_P}\Big)^2\Big\{\frac{2|y_t|^2}{8\pi^2}\Big(\frac{m_{\tilde{\Psi}}}{M_P}\Big)^2+\Big(\frac{M}{M_P}\Big)^2\Big(1-\frac{4M^2}{m^2_{\tilde{\Psi}}}\Big)^{\frac{1}{2}}\Big\}\,.
 \label{gr21}
\end{eqnarray}
Then, from Eqs.\,(\ref{gr11}) and (\ref{gr21}) the decay rate of inflaton through gravitational egffects is much smaller than that of the waterfall field, {\it i.e.} $\Gamma^{{\rm sugra}}_{\tilde{\Psi}}\gg\Gamma^{{\rm sugra}}_{\Psi_0}$, for $m_{\Psi_0}\gg m_{3/2}$. And  the waterfall field decay rate through the gravitational coupling of the waterfall field to matter is approximately given by
\begin{eqnarray}
\Gamma^{{\rm sugra}}_{\tilde{\Psi}}&\simeq&\Gamma^{t({\rm sugra})}_{\tilde{\Psi}}+\Gamma^{N({\rm sugra})}_{\tilde{\Psi}}=\frac{|g^{\rm sugra}_{\tilde{\Psi}}|^2}{8\pi}m_{\tilde{\Psi}}\,,
 \label{sgrNt}
\end{eqnarray}
where
\begin{eqnarray}
g^{\rm sugra}_{\tilde{\Psi}}\equiv\frac{\mu_\Psi(t_I)}{M_P}\Big\{\frac{|y_t|^2}{8\pi^2}\Big(\frac{m_{\tilde{\Psi}}}{M_P}\Big)^2+\frac{1}{2}\Big(\frac{M}{M_P}\Big)^2\Big(1-\frac{4M^2}{m^2_{\tilde{\Psi}}}\Big)^{\frac{1}{2}}\Big\}^{\frac{1}{2}}\,.
 \label{sgrNt1}
\end{eqnarray}
Given that $m_{\tilde{\Psi}}\sim10^{13}$ GeV, $\mu_\Psi(t_I)\sim10^{16}$ GeV, $M\sim10^9$ GeV, $y_t\sim1$, and $m_{3/2}\sim{\cal O}(100)$ TeV, we clearly have $\Gamma^{\rm vis}_{\Psi_0}+\Gamma^{{\rm sugra}}_{\tilde{\Psi}}\gg\Gamma^{{\rm sugra}}_{\Psi_0}+\Gamma^{\rm vis}_{\tilde{\Psi}}$ for $g_{\Psi_0}\sim g^{\rm sugra}_{\tilde{\Psi}}$, and
\begin{eqnarray}
g^{\rm sugra}_{\tilde{\Psi}}\sim10^{-9}\,.
 \label{gsug}
\end{eqnarray}
Then the total decay rate of the inflaton and waterfall fields in Eq.\,(\ref{tot_dec}) is approximately given by
 \begin{eqnarray}
 \Gamma_{\rm all}&\simeq&\Gamma^{\rm vis}_{\Psi_0}+\Gamma^{\rm sugra}_{\tilde{\Psi}}
 \label{gaPsi}
\end{eqnarray}
which is much larger than $\Gamma_{3/2}$ in Eq.\,(\ref{ga32}).
Putting Eqs.\,(\ref{dec_psi}) and (\ref{sgrNt}) into Eq.\,(\ref{inf_reht}), the reheating temperature can be expressed as
   \begin{eqnarray}
 T_{\rm reh}\simeq\left(\frac{10}{\pi^2 g_\ast}\right)^{1/4}\sqrt{m_{\Psi_0}M_{P}(|g_{\Psi_0}|^2+|g^{\rm sugra}_{\tilde{\Psi}}|^2)}\,,
  \label{inf_reht1}
 \end{eqnarray}
 where $m_{\tilde{\Psi}}\simeq m_{\Psi_0}$ is used.
 Since there is no information on the size of the renormalizable superpotential coupling $g_{\Psi_0}$ of the inflaton to the Higgses and Higgssinos, 
 first we consider the case of $\Gamma_{\rm all}\simeq\Gamma^{\rm vis}_{\Psi_0}\gg\Gamma^{\rm vis}_{\tilde{\Psi}}+\Gamma^{\rm sugra}_{\tilde{\Psi}}+\Gamma^{\rm sugra}_{\Psi_0}$. In this case, that is, $g_{\Psi_0}\gg|g^{\rm sugra}_{\tilde{\Psi}}|$, the size of the Higgs-inflaton coupling can severely restrict the lower limit on $T_{\rm reh}$ in Eq.\,(\ref{inf_reht1}) as
\begin{eqnarray}
 T_{\rm reh}\gtrsim10^4\,{\rm TeV}\left(\frac{g_{\Psi_0}}{10^{-8}}\right)\left(\frac{\tilde{g}_7}{0.94\times10^{-3}}\right)^{1/2}\left(\frac{\mu_\Psi(t_I)}{6.7\times10^{15}\,{\rm GeV}}\right)^{1/2}
 \label{lowlimit}
\end{eqnarray}
where we have used $m_{\Psi_0}=|\tilde{g}_7|\,\mu_\Psi(t_I)$ in Eqs.\,(\ref{inf_ma}) and (\ref{infl_ma1}). This lower limit\,\footnote{Note that, as seen from Fig.\,\ref{Fig2}, for values of $\delta_{\rm eff}$ being fine-tuned, {\it i.e.} $\delta_{\rm eff}<0.01$, the lower limit Eq.\,(\ref{lowlimit}) could be allowed for a successful leptogenesis.} on $T_{\rm reh}$ is conflict with the limit for the successful leptogenesis in Eqs.\,(\ref{AD_6}) and (\ref{Tr_pre}) for $0.01\leq\delta_{\rm eff}\leq1$.
Hence we can conclude that for $|g^{\rm sugra}_{\tilde{\Psi}}|\gtrsim g_{\Psi_0}$ 
from Eq.\,(\ref{inf_reht1}) the reheating temperature is in a good approximation given in terms of Eq.\,(\ref{gsug}) by
\begin{eqnarray}
 T_{\rm reh}\sim10^3\,{\rm TeV}
  \label{let_reh}
\end{eqnarray}
for the successful letogenesis with Eqs.\,(\ref{AD_6}-\ref{Tr_pre}). 
Inserting Eqs.\,(\ref{ga32}) and (\ref{gaPsi}) into Eq.\,(\ref{y32t}), the production of the gravitinos can depend on the size of the Higgs-inflaton coupling
 \begin{eqnarray}
  Y^{\Psi_0}_{3/2}
  \simeq3.2\times10^{-17}\Big(\frac{8\times10^{-10}}{g_{\Psi_0}}\Big)^{2}\Big(\frac{T_{\rm reh}}{10^3\,{\rm TeV}}\Big)\Big(\frac{|\tilde{g}_7|}{0.94\times10^{-3}}\Big)^3\Big(\frac{\mu_{\Psi}(t_I)}{6.7\times10^{15}\,{\rm GeV}}\Big)^{5}\,.
  \label{toyi}
\end{eqnarray}
Since the yield $Y^{\Psi_0}_{3/2}$ is inversely proportional to $|g_{\Psi_0}|^2$ and proportional to $T_{\rm reh}$ ($Y^{\rm th}_{3/2}$ is also proportional to $T_{\rm reh}$), the total yield $Y_{3/2}\simeq Y^{\rm th}_{3/2}+Y^{\Psi_0}_{3/2}$ can depend on the size of the Higgs-inflaton coupling, $|g_{\Psi_0}|$, with the given reheating temperature for the successful leptogenesis. And the constraint $Y_{3/2}<Y^{\rm BBN}_{3/2}$ disappears as in Ref.\,\cite{Kohri:2005wn} for the gravitino mass $m_{3/2}\sim100$ TeV in Eq.\,(\ref{gravi32}) with the given reheating temperature.
So we have an upper bound on the size of the Higgs-inflaton coupling, $|g_{\Psi_0}|$, with the given reheating temperature for the successful leptogenesis;
 \begin{eqnarray}
  |g_{\Psi_0}|\lesssim|g^{\rm sugra}_{\tilde{\Psi}}|\simeq8\times10^{-10}\,.
  \label{toyi1}
\end{eqnarray}
Since the size of Higgs-inflaton coupling can have an upper bound with the given reheating temperature,
the first term in Eq.\,(\ref{nong}) can contribute to the sizable $\mu$-term.

\section{Conclusion}
The model is based on the $SM\times U(1)_X\times A_4$ symmetry, which is essential for the flavored PQ axions at low energy. Note that the $U(1)_X$-charged Kahler moduli superfields put the GS anomaly cancellation mechanism into practice. As the $U(1)_X$ breaking scales according to Ref.\,\cite{Ahn:2016hbn} are secluded by the Gibbons-Hawking temperature $T_{\rm GH}=H_I/2\pi$, the model is designed in a way that gravitational interactions explicitly break supersymmetry (SUSY) down to SUSY$_{\rm inf}\times$SUSY$_{\rm vis}$, where SUSY$_{\rm inf}$ corresponds to the supergravity symmetry, while the orthogonal SUSY$_{\rm vis}$ is approximate global symmetry. Hence, in the presence of SUGRA, the SUSY$_{\rm inf}$ is gauged and thus its corresponding goldstino is eaten by the gravitino via super-Higgs mechanism, leaving behind the approximate global symmetry SUSY$_{\rm vis}$ which is explicitly broken by SUGRA and thus its corresponding the uneaten goldstino as a physical degree of freedom giving masses to all the supersymmetric SM superpartners.

In order to provide an explanation for inflation we have considered a realistic supersymmetric moduli stabilization.
Such moduli stabilization has moduli backreaction effects on the inflationary potential, in particular, the spectral index of inflaton fluctuations. During inflation the universe experiences an approximately dS phase with the inflationary Hubble constant $H_I\simeq2\times10^{10}$ GeV.
In the present inflation model which provides intriguing links to UV-complete theories like string theory, the PQ scalar fields $\Psi(\tilde{\Psi})$ play a role of the waterfall fields, that is, the PQ phase transition takes place during inflation such that the PQ scale $\mu_\Psi(t_I)$ during inflation is fixed by the amplitude of the primordial curvature perturbation and turns out to be roughly $0.3\times10^{16}$ GeV.
We have found that such moduli stabilization with the moduli backreaction effects on the inflationary potential could lead to the energy scale of inflation in a way that the power spectrum of the curvature perturbation and the scalar spectral index are to be well fitted with the Planck 2015 observation\,\cite{Ade:2015xua}. And we have driven that the inflaton mass during inflation is given by $m_{\Psi_0}=\sqrt{3}\,H_I$ which is much larger than the gravitino mass, and its mass is in agreement with its theory prediction for spectral index with observation. 

Through the introduction of $U(1)_X$ symmetry in a way that the $U(1)_X$-$[gravity]^2$ anomaly-free condition together with the SM flavor structure demands additional sterile neutrinos as well as no axionic domain-wall problem\,\cite{Ahn:2016hbn}, the additional neutrinos may play a crucial role as a bridge between leptogenesis and new neutrino oscillations along with high energy cosmic events. 
We have shown that a successful leptogenesis scenario could be naturally implemented through Affleck-Dine mechanism.
The pseudo-Dirac mass splittings, which is suggested from new neutrino oscillations along with high energy cosmic events, strongly indicate the existence of lepton-number violation which is a crucial ingredient of the present leptogenesis scenario. The resultant baryon asymmetry is constrained by the cosmological observable ({\it i.e.} the sum of active neutrino masses) with the new high energy neutrino oscillations.
In addition, the resultant baryon asymmetry, which crucially depends on the reheating temperature, is suppressed for relatively high reheating temperatures. 
We have shown that the right value of BAU, $Y_{\Delta B}\simeq8\times10^{-11}$ prefers a relatively low reheating temperature with the well constrained pseudo-Dirac mass splittings responsible for new oscillations $\Delta m^2_i$. Moreover, we have shown that it is reasonable for the reheating temperature $T_{\rm reh}\sim10^3$ TeV derived from the gravitational decays of the inflaton and waterfall field to be compatible with the required reheating temperature for the successful leptogenesis, leading to $\Delta m^2_i\sim10^{-12}$ eV$^2$. We have stressed that the present model requires $m_{3/2}\simeq{\cal O}(100)$ TeV gravitino mass in order to have suitable large gaugino masses.

\newpage
\appendix
\section{Superpotential dependent on driving fields}
\label{dri}
To impose the $A_{4}$ flavor symmetry\,\cite{Ma:2001dn} on our model properly, apart from the usual two Higgs doublets $H_{u,d}$ responsible for electroweak symmetry breaking, which are invariant under $A_{4}$ ({\it i.e.} flavor singlets $\mathbf{1}$ with no $T$-flavor), the scalar sector is extended by introducing two types of new scalar multiplets, flavon fields $\Phi_{T},\Phi_{S},\Theta,\tilde{\Theta}, \Psi, \tilde{\Psi}$ that are $SU(2)$-singlets and driving fields $\Phi^{T}_{0},\Phi^S_{0},\Theta_{0},\Psi_{0}$ that are associated to a nontrivial scalar potential in the symmetry breaking sector: we take the flavon fields $\Phi_{T},\Phi_{S}$ to be $A_{4}$ triplets, and $\Theta,\tilde{\Theta},\Psi,\tilde{\Psi}$ to be $A_{4}$ singlets with no $T$-flavor ($\mathbf{1}$ representation), respectively, that are $SU(2)$-singlets, and driving fields $\Phi_{0}^{T},\Phi_{0}^{S}$ to be $A_{4}$ triplets and $\Theta_{0}, \Psi_{0}$ to be an $A_{4}$ singlet.
Under $A_{4}\times U(1)_{X}\times U(1)_{R}$, the driving, flavon, and Higgs fields are assigned as in TABLE\,\ref{DrivingRef}.
\begin{table}[h]
\caption{\label{DrivingRef} Representations of the driving, flavon, and Higgs fields under $A_4 \times U(1)_{X}$. Here $U(1)_X\equiv U(1)_{X_1}\times U(1)_{X_2}$ symmetries which are generated by the charges $X_1=-2p$ and $X_2=-q$.}
\begin{ruledtabular}
\begin{tabular}{cccccccccccccccc}
Field &$\Phi^{T}_{0}$&$\Phi^{S}_{0}$&$\Theta_{0}$&$\Psi_{0}$&\vline\vline&$\Phi_{S}$&$\Phi_{T}$&$\Theta$&$\tilde{\Theta}$&$\Psi$&$\tilde{\Psi}$&\vline\vline&$H_{d}$&$H_{u}$\\
\hline
$A_4$&$\mathbf{3}$&$\mathbf{3}$&$\mathbf{1}$&$\mathbf{1}$&\vline\vline&$\mathbf{3}$&$\mathbf{3}$&$\mathbf{1}$&$\mathbf{1}$&$\mathbf{1}$&$\mathbf{1}$&\vline\vline&$\mathbf{1}$&$\mathbf{1}$\\
$U(1)_{X}$&$0$&$4p$&$4p$&$0$&\vline\vline&$-2p$&$0$&$-2p$&$-2p$&$-q$&$q$&\vline\vline&$0$&$0$\\
$U(1)_R$&$2$&$2$&$2$&$2$&\vline\vline&$0$&$0$&$0$&$0$&$0$&$0$&\vline\vline&$0$&$0$\\
\end{tabular}
\end{ruledtabular}
\end{table}
The superpotential dependent on the driving fields, which is invariant under  $SU(3)_c\times SU(2)_L\times U(1)_{Y}\times U(1)_{X}\times A_{4}$, is given at leading order by
 \begin{eqnarray}
W_{v} &=& \Phi^{T}_{0}\left(\tilde{\mu}\,\Phi_{T}+\tilde{g}\,\Phi_{T}\Phi_{T}\right)+\Phi^{S}_{0}\left(g_{1}\,\Phi_{S}\Phi_{S}+g_{2}\,\tilde{\Theta}\Phi_{S}\right)\nonumber\\
 &+& \Theta_{0}\left(g_{3}\,\Phi_{S}\Phi_{S}+g_{4}\,\Theta\Theta+g_{5}\,\Theta\tilde{\Theta}+g_{6}\,\tilde{\Theta}\tilde{\Theta}\right)+g_{7}\,\Psi_{0}\left(\Psi\tilde{\Psi}-\mu^2_\Psi\right)\,,
 \label{potential}
 \end{eqnarray}
where the fields $\Psi$ and $\tilde{\Psi}$ charged by $-q,q$, respectively, are ensured by the $U(1)_{X}$ symmetry extended to a complex $U(1)$ due to the holomorphy of the supepotential. SUSY hybrid inflation, defined by the last term in the above superpotential, provides a compelling framework for the understanding of the early universe, where $\Psi_0$ and $\Psi(\tilde{\Psi})$ are identified as the inflaton and waterfall fields, respectively. 
Note here that the PQ scale $\mu_\Psi\equiv\sqrt{v_{\Psi}v_{\tilde{\Psi}}/2}$ corresponding to the scale of the spontaneous symmetry breaking scale sets the energy scale of inflation during inflation, see Eq.\,(\ref{PQ_scale2}), as well as the energy scale at present in Ref.\,\cite{Ahn:2016hbn}.

\section{A direct link between Low and High energy Neutrinos}
\label{low_nut}
Once the scalar fields $\Phi_{S}, \Theta, \tilde{\Theta},\Psi$ and $\tilde{\Psi}$ get VEVs, the flavor symmetry $U(1)_{X}\times A_{4}$ is spontaneously broken
And at energies below the electroweak scale, all leptons obtain masses.
Since the masses of Majorana neutrino $N_R$ are much larger than those of Dirac and light Majorana ones, after integrating out the heavy Majorana neutrinos, we obtain the following effective Lagrangian for neutrinos
 \begin{eqnarray}
  -{\cal L}^{\nu}_{W} &\simeq&\frac{1}{2} \begin{pmatrix} \overline{\nu^{c}_L} & \overline{S_R} \end{pmatrix} {\cal M}_{\nu} \begin{pmatrix} \nu_L \\ S^{c}_R \end{pmatrix}+\frac{1}{2}\overline{N_R}\,M_R\,N^c_R + \overline{\ell_{R}}\,{\cal M}_{\ell}\,\ell_{L}+\frac{g}{\sqrt{2}}W^-_\mu\overline{\ell_{L}}\gamma^\mu\,\nu_{L}+\text{h.c.}
\label{Axion_nu_La}\\
\text{with}&&~~{\cal M}_{\nu}= \begin{pmatrix} -m^T_DM^{-1}_Rm_D & m^T_{DS}  \\ m_{DS} &  M_{S}   \end{pmatrix}\,.
  \label{neut1}
 \end{eqnarray}
And the charged lepton mass term and the Dirac and Majorana neutrino mass terms read 
 \begin{eqnarray}
 {\cal M}_{\ell}&=& {\left(\begin{array}{ccc}
 y_{e} & 0 &  0 \\
 0 & y_{\mu} & 0 \\
 0 & 0 & y_{\tau}
 \end{array}\right)}v_{d}= {\left(\begin{array}{ccc}
 (\frac{\lambda}{\sqrt{2}})^4\,\hat{y}_{e}& 0 &  0 \\
 0 & (\frac{\lambda}{\sqrt{2}})^2\,\hat{y}_{\mu} & 0 \\
 0 & 0 & \hat{y}_{\tau}
 \end{array}\right)}\left(\frac{\lambda}{\sqrt{2}}\right)^2v_{d}\,,
 \label{ChL3}\\
m_{DS}&=&{\left(\begin{array}{ccc}
 \hat{y}^s_{1} &  0 &  0 \\
 0 &  \hat{y}^s_{2} &  0   \\
 0 &  0  &  \hat{y}^s_{3}
 \end{array}\right)}\left(\frac{v_{\Psi}}{\sqrt{2}\Lambda}\right)^{16}v_{u}, \label{YDS1}\\
 M_{S}&=&{\left(\begin{array}{ccc}
 \hat{y}^{ss}_{1} &  0 &  0 \\
 0 &  0 &  \hat{y}^{ss}_{2} \\
 0 &  \hat{y}^{ss}_{2} &  0 \end{array}\right)}\frac{v_{\tilde{\Psi}}}{\sqrt{2}}\left(\frac{v_{\Psi}}{\sqrt{2}\Lambda}\right)^{51}\frac{v_\Theta}{\sqrt{2}\Lambda}\,,\label{YS1}
  \end{eqnarray}
  \begin{eqnarray}
 m_{D}&=&{\left(\begin{array}{ccc}
 \hat{y}^{\nu}_{1} &  0 &  0 \\
 0 &  0 &  \hat{y}^{\nu}_{2}   \\
 0 &  \hat{y}^{\nu}_{3}  &  0
 \end{array}\right)}\frac{v_{T}}{\sqrt{2}\Lambda}\left(\frac{v_{\tilde{\Psi}}}{\sqrt{2}\Lambda}\right)^{9}v_{u}
 =\hat{y}^{\nu}_{1}{\left(\begin{array}{ccc}
 1 &  0 &  0 \\
 0 &  0 &  y_{2}   \\
 0 &  y_{3}  &  0
 \end{array}\right)}\frac{v_{T}}{\sqrt{2}\Lambda}\left(\frac{v_{\tilde{\Psi}}}{\sqrt{2}\Lambda}\right)^{9}v_{u}, \label{Ynu1}\\
 M_{R}&=&{\left(\begin{array}{ccc}
 1+\frac{2}{3}\tilde{\kappa}\,e^{i\phi} &  -\frac{1}{3}\tilde{\kappa}\,e^{i\phi} &  -\frac{1}{3}\tilde{\kappa}\,e^{i\phi} \\
 -\frac{1}{3}\tilde{\kappa}\,e^{i\phi} &  \frac{2}{3}\tilde{\kappa}\,e^{i\phi} &  1-\frac{1}{3}\tilde{\kappa}\,e^{i\phi}\\
 -\frac{1}{3}\tilde{\kappa}\,e^{i\phi} &  1-\frac{1}{3}\tilde{\kappa}\,e^{i\phi} &  \frac{2}{3}\tilde{\kappa}\,e^{i\phi}
 \end{array}\right)}M~,
 \label{MR1}
 \end{eqnarray}
where $v_{d}\equiv\langle H_{d}\rangle=v\cos\beta/\sqrt{2}$, and $v_{u}\equiv\langle H_{u}\rangle =v\sin\beta/\sqrt{2}$ with $v\simeq246$ GeV, and
 \begin{eqnarray}
 y_{2}&\equiv&\frac{\hat{y}^{\nu}_{2}}{\hat{y}^{\nu}_{1}}~,\quad y_{3}\equiv\frac{\hat{y}^{\nu}_{3}}{\hat{y}^{\nu}_{1}}~,\quad\tilde{\kappa}\equiv\sqrt{\frac{3}{2}}\left|\hat{y}_{R}\frac{v_{S}}{M}\right|~,\quad\phi\equiv\arg\left(\frac{\hat{y}_{R}}{\hat{y}_{\Theta}}\right)~\,\text{with}~M\equiv \left|\hat{y}_{\Theta}\,\frac{v_{\Theta}}{\sqrt{2}}\right|\,.
 \label{MR2}
 \end{eqnarray}
Here all the hat Yukawa couplings are of order unity.

In Eq.\,(\ref{neut1}) the Majorana neutrino mass terms $M_{\nu\nu}$ and $M_{S}$, and the Dirac mass term $m_{DS}$ are given by
 \begin{eqnarray}
  M_{\nu\nu}=U^{\ast}_L\hat{M}_{\nu\nu}U^\dag_L=-m^T_DM^{-1}_Rm_D \,,\qquad M_{S}=U^\ast_R\hat{M}_{S}U^\dag_R\,,\qquad
  m_{DS}=U^\ast_R\,\hat{M}\,U^\dag_L\,,
 \end{eqnarray}
where ``hat" matrices represent diagonal mass matrices of their corresponding leptons, and $U_{L(R)}$ are their diagonal left(right)-mixing matrix. 
Since $m_{DS}$ is dominant over $M_{\nu\nu}$ and $M_S$ due to Eqs.\,(\ref{YDS1}-\ref{MR1}), the low energy effective light neutrinos become pseudo-Dirac particles.

The pseudo-Dirac mass splitting, $\delta$, can be given by
 \begin{eqnarray}
   \delta\equiv\hat{M}_{\nu\nu}+\hat{M}^\dag_S\simeq\hat{M}_{\nu\nu}\,,
   \label{m_split0}
 \end{eqnarray}
where the second equality is due to $|\hat{M}_{\nu\nu}|\gg|\hat{M}_S|$. 
As is well-known, because of the observed hierarchy $|\Delta m^{2}_{\rm Atm}|= |m^{2}_{\nu_3}-(m^{2}_{\nu_1}+m^{2}_{\nu_2})/2|\gg\Delta m^{2}_{\rm Sol}\equiv m^{2}_{\nu_2}-m^{2}_{\nu_1}>0$, and the requirement of a Mikheyev-Smirnov-Wolfenstein resonance for solar neutrinos, there are two possible neutrino mass spectra: (i) the normal mass ordering (NO) $m^2_{\nu_1}<m^2_{\nu_2}<m^2_{\nu_3},~m^2_{s_1}<m^2_{s_2}<m^2_{s_3}$, and (ii) the inverted mass ordering (IO) $m^2_{\nu_3}<m^2_{\nu_1}<m^2_{\nu_2},~m^2_{s_3}<m^2_{s_1}<m^2_{s_2}$, in which the mass-squared differences in the $k$-th pair $\Delta m^2_{k}\equiv m^2_{\nu_k}-m^2_{s_k}$ are enough small that the same mass ordering applies for the both eigenmasses, that is, 
 \begin{eqnarray}
   \Delta m^2_{k}=2m_k|\delta_k|\ll m^2_{\nu_k}
  \label{msd}
 \end{eqnarray}
 for all $k=1,2,3$.
It is anticipated that $\Delta m^2_k\ll\Delta m^2_{\rm Sol}, |\Delta m^2_{\rm Atm}|$,
otherwise the effects of the pseudo-Dirac neutrinos should have been detected.
But in the limit that $\Delta m^2_k=0$, it is hard to discern the pseudo-Dirac nature of neutrinos. 
The pseudo-Dirac mass splittings could be limited by several constraints, that is, the active neutrino mass hierarchy, the BBN constraints on the effective  number of species of light particles during nucleosynthesis, the solar neutrino oscillations:
we roughly estimate a bound for the tiny mass splittings
 \begin{eqnarray}
   6\times10^{-16}\lesssim\Delta m^2_k/{\rm eV}^2\lesssim1.8\times10^{-12}\,,
  \label{D_lbound1}
 \end{eqnarray}
where the upper bound comes form the solar neutrino oscillations\,\cite{deGouvea:2009fp}, and the lower bound comes from the inflationary (Sec.\,\ref{inflat}) and leptogenesis (Sec.\,\ref{AD}) scenarios by assuming\,\footnote{In the present model the lightest effective neutrino mass could not be extremely small because the values of $\delta_k$ through the relation Eq.\,(\ref{msd}), are constrained by the $\mu-\tau$ powered mass matrix in Eq.\,(\ref{mass matrix}).} $m_{\nu_i}\sim0.01$ eV.

Letting the mass of active neutrino $m_{\nu_k}=m_k$, then
 the sum of light neutrino masses given by
 \begin{eqnarray}
 \sum_{k}m_{\nu_k}=\frac{1}{2}\left(\frac{\Delta m^2_1}{\delta_1}+\frac{\Delta m^2_2}{\delta_2}+\frac{\Delta m^2_3}{\delta_3}\right)
  \label{nu_sum_c}
 \end{eqnarray}
is bounded by $0.06\lesssim\sum_{i}m_{\nu_i}/\text{eV}<0.194$;
the lower limit is extracted from the neutrino oscillation measurements, and 
the upper limit\,\footnote{Massive neutrinos could leave distinct signatures on the CMB and large-scale structure at different epochs of the universe's evolution\,\cite{Abazajian:2008wr}. To a large extent, these signatures could be extracted from the available cosmological observations, from which the total neutrino mass could be constrained. } is given by Planck Collaboration\,\cite{Planck2014} which is subject to the cosmological bounds $\sum_{i}m_{\nu_i}<0.194$ eV at $95\%$ CL (the CMB temperature and polarization power spectrum from Planck 2015
in combination with the BAO data, assuming a standard $\Lambda$CDM cosmological model).

\acknowledgments{We would like to give thanks to two anonymous referees for very useful comments, and Eibun Senaha and MH Ahn for useful conversations. The work is supported in part by Basic Science Research Program through the National Research Foundation of Korea (NRF) funded by the Ministry of Education, Science and Technology (NRF-2019R1A2C2003738).
}


\end{document}